\documentclass[12pt]{article}
\usepackage{float}

\usepackage{amssymb, amsmath, stmaryrd, footmisc}
\usepackage{color}
\usepackage{siunitx}
\usepackage{mathtools}
\usepackage[toc, page]{appendix}

\def\be{\begin{equation}}
\def\ee{\end{equation}}
\def\ba{\begin{array}}
\def\ea{\end{array}}

\usepackage{geometry}                
\usepackage{fullpage}
\usepackage{amsmath,amsfonts,mathrsfs,array,pifont,mathrsfs}
\usepackage{amsthm}
\usepackage{mathrsfs}
\usepackage{graphicx}
\usepackage{graphics}
\usepackage{amssymb}
\usepackage{epstopdf}
\usepackage{epsfig}
\usepackage{times}
\usepackage{subfigure}
\usepackage{hyperref}
\usepackage{color}
\usepackage{yfonts}
\usepackage{cancel}
\usepackage{latexsym}
\usepackage{cite}
\usepackage{multirow}
\usepackage{url}
\usepackage{algorithmic}
\usepackage{mathtools}
\usepackage{stmaryrd}
\usepackage{bm}

\begin{document}


\title{
A continuum model for the\\[1ex]
 growth of dendritic actin networks\\
}

\begin{center}
{\Large A continuum model for the\\[1ex]
 growth of dendritic actin networks}\\[4ex]

Rohan Abeyaratne$^{1}$ and Prashant K. Purohit$^{2}$\\[2ex]

$^{1}$Department of Mechanical Engineering,\\ Massachusetts Institute of Technology,\\
 Cambridge,  Massachusetts, 02139, USA\\[1ex]
 
$^{2}$Department of Mechanical Engineering and Applied Mechanics,\\ University of Pennsylvania,\\  Philadelphia, Pennsylvania, 19104, USA\\[4ex]

June 15, 2020\\
\end{center}

\begin{abstract}
Polymerization of dendritic actin networks underlies important mechanical processes in 
cell biology such as the protrusion of lamellipodia, propulsion of growth cones 
in dendrites of neurons, intracellular transport of organelles and pathogens, 
among others. The forces required for these mechanical functions have been deduced from
mechano-chemical models of actin polymerization; most models are focused on single 
growing filaments, and only
a few address polymerization of filament networks through simulations. Here we propose
a continuum model of surface growth and filament nucleation to describe polymerization of dendritic actin
networks. The model describes growth and elasticity in terms of macroscopic stresses,
strains and filament density rather than focusing on individual filaments. The
microscopic processes underlying polymerization are subsumed into kinetic laws
characterizing the change of filament density and the propagation of growing surfaces. This continuum 
model can predict the evolution of actin networks in disparate experiments. A key conclusion of the analysis is that 
existing laws relating force to polymerization speed of single filaments cannot
predict the response of growing networks. Therefore a new kinetic law, consistent with the dissipation inequality, is proposed
to capture the evolution of dendritic actin networks under different loading
conditions. This model may be
extended to other settings involving a more complex interplay between mechanical stresses 
and polymerization kinetics, such as the growth of networks of microtubules, collagen
filaments, intermediate filaments and carbon nanotubes.

\end{abstract}



\section{Introduction.} \label{sec-1}

Actin polymerization drives the protrusion of lamellipodia and pseudopodia that
play important roles in cell sensing and motility~\cite{boal,howardbook,
phillips}. Actin polymerization is also
the mechanism behind motion of growth cones in dendrites of neurons, 
organelles in cells, as well as pathogens, such as, {\it Listeria} and {\it Shigella}~\cite{theriot}. 
As such actin based motility has been studied using experiment and theory for 
at least the last three decades~\cite{howardbook}. It is now understood that the propulsive 
force in actin polymerization is a result of the difference in chemical 
potential of the actin monomers in their bound (to the filament) and unbound 
(in solution) states~\cite{brangbour,theriot}.

Early Brownian ratchet models \cite{peskin} for the force generated by actin 
polymerization focused on a single
filament polymerizing against a load represented by a bead moving through 
a viscous medium, and the goal was to relate the force on the filament with its
elongation-rate measured by the
velocity of the bead~\cite{howardbook, philnel}. In these models thermal 
fluctuations of the filament and the bead created gaps between the bead and 
the polymerizing tip and resulted in an exponential dependence of the 
velocity $V$ of the bead and the 
force $f$ on the filament (taken to be positive in compression):
\begin{equation} \label{eq:expokin}
  V = V_{0}\frac{\exp(-\zeta {f}/{f_{stall}}) - \exp(-\zeta)}{1 - \exp(-\zeta)},
\end{equation}
where $V_{0}$ is the velocity at $f = 0$ and $f_{stall}$ is the force at which
polymerization stalls~\cite{philnel}; $\zeta$ is related to the size of the 
monomers $a$ through $\zeta = {a f_{stall}}/{(k_{B}T)}$ where $k_{B}$ is the Boltzmann
constant and $T$ the absolute temperature. The stall force for actin was 
estimated to be around $1 - 8 \, \si{pN}$ \cite{howardbook}
and this was confirmed through experiments on growing bundles of a few actin filaments 
with forces exerted using an optical 
trap~\cite{theriot}. However, dendritic actin and the actin gel in {\it Listeria} `comet 
tails' is not a bundle of parallel growing filaments; rather, it is branched 
and cross-linked with filament lengths that are in the range of 1 $\mu$m or 
less~\cite{brangbour,marcy,ParekhChaudhuri2005}. Thus, in dendritic actin a large number of short filaments exert 
forces on the lamellipodial membrane or the moving object (pathogen, bead, 
organelle, etc.) resulting in an elongation rate-force\footnote{We will soon distinguish 
elongation-rate from growth-rate, but we note that they are the same when the monomers do not deform.} 
curve that is not necessarily
of the form (\ref{eq:expokin}). This was shown in subsequent models 
which accounted for the transient nature of the contact between the load 
surface and growing filaments and the elasticity of the filament 
network~\cite{prostgel,mogiloster}. 
These models and others~\cite{leeliu} recognized that polymerization 
in dendritic actin occurs in a narrow
zone next to the load surface (not all over the network), so that the barbed ends 
of most of the growing actin filaments are pointed toward the load surface.

Experiments to measure the elongation rate-force relation of dendritic actin are
scarce. One experiment that yielded intriguing results was that of 
Parekh {\it et al.}~\cite{ParekhChaudhuri2005} who polymerized dendritic actin between two atomic 
force microscope (AFM) cantilevers. The deflection of the AFMs allowed them 
to accurately measure both forces and elongation-rates. They were
also able to control the conditions under which polymerization occurred (e.g.,
constant force, a force proportional to network height, etc.). They found that
the elongation-rate was independent of the force over a range of force and 
this was followed by a {\it convex} curve in which polymerization stalled over
a short range of force; this elongation rate-force relation is very different from
the exponential form in equation (\ref{eq:expokin}). Parekh {\it et al.} also
demonstrated that the elongation-rate was loading history dependent and in particular that 
there could be two (or more) steady state elongation-rates at the same force. 
Some of these findings were confirmed in a later experiment by 
Brangbour {\it et al.}~\cite{brangbour} who used magnetic beads to control the force resisting 
polymerization. The actin polymerized between two beads with short filaments contacting 
the beads at random angles, much like a dendritic 
network. Polymerization caused the beads to move away from each other so that 
their velocity could be measured as a function of the force
applied on them by the magnetic field. These authors showed that constraints 
on the rotational fluctuations of the filaments resulted in an elongation rate-force 
relation that is different from the exponential form of equation (\ref{eq:expokin}).

Our objective in this paper is to explain the results of Brangbour {\it et al.}  \cite{brangbour}
and Parekh {\it et al.} \cite{ParekhChaudhuri2005} through a continuum model of growth. Continuum {\it models}\footnote{There is a vast literature on this topic which we do not attempt to review here. Some books and review articles that the interested reader can refer to include \cite{ArrudaKuhlGarikipati2011, garikipati, Goriely:2017, jones, Kuhl2014, kuhl2012, Taber1995, epstein, ProstJuelicher2015, Zimmermann2014, BindschadlerOsborn2004, CardamoneLaio2011}}  of growth fall into two broad categories: volumetric growth and surface growth. In the former, new material is added to existing material points whereas in the latter, {\it new material points} are added to the body at its evolving boundary.  One way in which to track the newly added material points is through a continuously evolving reference configuration\footnote{The seminal paper by Skalak {\it et al.} \cite{SkalakDasgupta1982} appears to be the first to model the kinematics of this.  For a complete treatment of the kinematics, mechanics, thermodynamics and kinetics of surface growth, see, e.g. \cite{GTTCRA2016}.}.  The driving force for surface growth is essentially the configurational force associated with this evolving reference configuration and can be calculated in several different\footnote{See for example \cite{Ganghoffer2010, Ganghoffer2018, BacigalupoGambarotta:2012, TruskinovskyZurlo2019, GTTCRA2016}} (roughly equivalent) ways.  The bio-physical micro-mechanisms underlying the growth process are captured at the continuum scale by a kinetic law for growth, with the second law of thermodynamics imposing certain restrictions on such a kinetic law.

In our continuum model the body is treated as a one-dimensional nonlinearly elastic bar whose length at time $t$ is $\ell(t)$ in physical space and $\ell_R(t)$ in reference space. The body we are concerned with is comprised of actin filaments (F-actin) -- long polymer chains -- formed by the assembly of actin monomers (G-actin). It is surrounded by a pool of free actin monomers that, under suitable conditions, 
can bind to the tips of the existing filaments leading to their growth. The length $\ell_R(t)$  evolves due to the addition of new material points. In addition, the formation of new filaments leads to a 
time-dependent filament density $\rho(t)$ \cite{ParekhChaudhuri2005}. At this juncture, it is worth emphasizing the distinction between the elongation-rate $\dot \ell$ and the growth-rate $\dot\ell_R$.  The former is the rate of change of the length of the specimen in physical space, a quantity that can be observed and measured. Figure \ref{Fig-elldot-sig2.pdf} shows a plot of $\dot\ell$ versus the force $\sigma A$ as predicted by our model in a particular setting; note the commonly observed rapidly declining nature of this curve.  On the other hand, any change in the length $\ell_R$ in reference space occurs solely due to growth. This is in contrast to $\ell$ which changes due to both growth and stress. The relation between $\dot\ell_R$ and force is an input into the model, the kinetic law for growth. The particular relation used to predict Figure \ref{Fig-elldot-sig2.pdf} is shown dotted in Figure \ref{Fig-KineticLaws}, and the qualitative distinction between the curves in the two figures is worth noting.

A phenomenon that brings out the distinction between growth and elongation is `treadmilling', commonly observed in growing actin filaments \cite{howardbook, boal}. During treadmilling, the filament length $\ell$ (in physical space) remains constant because of a precise balance between the rates at which monomers attach to one end (growth) of a filament and detach from the other. Thus, growth occurs
continuously, but elongation does not.  A 
recent paper \cite{ra2020} analyzes the existence and stability of treadmilling solutions in a one-dimensional elastic bar attached to a spring at one end. In contrast, in the setting studied in the present paper, monomers do not detach from either end (actin tends not to depolymerize at physiological ion concentrations and in the presence of capping proteins), so treadmilling does not
occur. Despite this, the length of the specimen can stop changing due to the phenomenon of {\it stall} which occurs when the compressive force is so large that it prevents the addition of new monomers
\cite{howardbook,boal}.

In the rest of this section we describe (with little justification) some key aspects of our model. These will be explained in detail in later sections. Let $\sigma$ and $\lambda$ denote the compressive stress and stretch in the body. Since the material stiffens with increasing compressive stress, we take a simple stress-stretch relation that captures this phenomenon: $\sigma = E(\lambda^{-1}-1)$.  The tangent modulus of the material, $-\partial \sigma/\partial \lambda$, then increases with increasing  $\sigma$. The elastic modulus $E$ could also depend on the filament density $\rho$. We are concerned with two settings: in the first (pertaining to Brangbour {\it et al.} \cite{brangbour}),  the number of actin filaments is relatively small and therefore do not form a cross-linked network. Consequently, filament bending is unimportant and so we take the elastic modulus to depend linearly on the filament density: $E(\rho) \sim \rho$. In the second setting (pertaining to Parekh {\it et al.} \cite{ParekhChaudhuri2005}) the filaments form a cross-linked network where filament bending plays an important role. Therefore in this setting we take the modulus to depend quadratically on the filament density: $E(\rho) \sim \rho^2$.  Moreover, in the first setting the filament density remains constant, $\dot \rho = 0$, whereas in the second, it changes over time which we model through a suitable evolution law $\dot \rho = \mathsf{R}(\rho)$. In both settings, new material points can be added to one end of the specimen leading to surface growth through polymerization. The growing boundary propagates at a speed $V$  given by a kinetic law
$V = \mathsf{V}(f)$ where $f=\sigma/\rho$ is the filament force. Finally, in the first setting, both ends of the specimen are attached to supports and the length or force on the specimen can be controlled. In the second setting, one end is attached to an AFM cantilever modeled as a Hookean spring so that, in addition to controlling the force, it is also possible to allow the specimen to freely evolve against the spring.  Because of the way in which the experiments are set-up,  polymerization only occurs at one of the end of the specimen.  At the other end, neither polymerization nor depolymerization takes place.

This paper is organized as follows: in Section \ref{sec-2} we describe the basic aspects of the model that are common to both settings studied in the subsequent sections.  Section \ref{sec-3} is concerned with problems where the filament density is time independent. Further details of the model are given in Section \ref{subsec-3-1}; the response of the model to various loading programs are examined in Section \ref{subsec-3-2} and compared with the experimental measurements of Brangbour {\it et al.}  \cite{brangbour}; and finally in Section \ref{sec-brangmodel} we explain how our model describes the setting in  \cite{brangbour} even though they might seem different at first glance. We then turn in Section \ref{sec-4} to the growth of a network of actin filaments.  The detailed constitutive model is presented in Section \ref{sec:20191220-1} and the values of the various parameters are given in Section \ref{subsubsec-parametervalues}. In Section \ref{subsec-spring} we study the response of the specimen when it is growing under the action of the spring, and in Section \ref{sec-0505-13} we examine its response to loading programs that involve a sudden change in the force. The results of both sections are compared with the experimental measurements of Parekh {\it et al.} \cite{ParekhChaudhuri2005}. In the appendix we derive the dissipation inequality, identify the driving force, and examine the kinetic laws considered in Sections  \ref{sec-3} and \ref{sec-4} in the context of the dissipation inequality. The Supplementary Material provides details on how the parameter values were chosen, the calculations in Section \ref{sec-0505-13} were carried out, and the conformity of the solutions in Sections \ref{sec-3} and \ref{sec-4} with the dissipation inequality.



\section{The basic model.} \label{sec-2}

Since we will examine two rather different experimental settings in this paper, here we simply write down those constituents of the mathematical model that are common to both.   More detailed descriptions will be given in Sections \ref{subsec-3-1} and  \ref{sec:20191220-1}.

Imagine a  test specimen composed of actin filaments held between two supports. One of the supports will be compliant in Section \ref{sec-4} as shown schematically in Figure \ref{Fig-TestSpecimen.pdf}. Each filament is a polymeric molecule, comprised of a linear assembly of monomers.  The filaments are surrounded by a solvent containing free monomers that can attach to one end  of each filament (called the barbed 
end in actin).

In physical space the specimen is identified with the interval $[y_0(t), y_1(t)]$ at time $t$ and so its corresponding length  is 
$$
\ell(t) = y_1(t) - y_0(t). 
$$
In reference space it is identified (at the same instant $t$) with the interval $[x_0(t), x_1]$ where $x_0$ is permitted to be a function of time because growth may occur at that end. No growth occurs at\footnote{Growth only occurs at one end of the specimen in the experiments of Parekh {\it et al.} \cite{ParekhChaudhuri2005} because  the support at that end was functionalized with an actin nucleating agent.}  $x_1$.  The length of the specimen in reference space at time $t$ is
\be
\label{eq-2xx}
\ell_{\rm R}(t) = x_1 - x_0(t). 
\ee
The length in physical space is affected by both stress and growth, whereas the length in reference space is only affected by growth.  When monomers are added to the specimen at its left-hand boundary, the specimen grows through the leftward motion of that surface in reference space at a speed
\be
\label{eq-2yy}
V = - \dot x_0.
\ee
Thus growth corresponds to $\dot x_0 <0$ (and therefore by \eqref{eq-2xx} to  $\dot \ell_{\rm R} >0$).  If there is no growth, $x_0(t)$ is constant.  In order not to confuse the two velocities $\dot\ell$ and $\dot\ell_R$, we shall refer to $\dot\ell$ as the {\it elongation-rate} and $\dot\ell_R$ as the {\it growth-rate}. In the problems of interest to us the stress and stretch fields are spatially uniform and so the stretch $\lambda$ of the specimen relates the reference and current lengths:
\be
\label{eq-3xx}
\ell = \lambda \, \ell_{\rm R}.
\ee

Let $N(t)$ denote the number of load bearing filaments in the specimen and let $A$ be the (fixed) area over which they are distributed. The filament density $\rho(t)$ is defined as the number of filaments per unit cross-sectional area:
$$
\rho \coloneqq N/A.  
$$
If the force in each filament is $f$, the total force in the specimen is $fN$ and so the stress is related to the filament force and filament density by
\be
\label{eq-0505-1}
\sigma =  fN/A = f \rho.
\ee
We are concerned exclusively with compressive stress and so take $\sigma$ to be positive in compression.

The key variables in the model are the stress $\sigma(t)$, stretch $\lambda(t)$, filament density $\rho(t)$, length of the specimen in physical space $\ell(t)$ and length of the specimen in reference space $\ell_R(t)$.  The stress depends on the stretch and the filament density through a constitutive relation $\sigma = \widehat \sigma (\lambda, \rho)$.  This relation is assumed to be invertible so that we can write
\be
\label{eq:20191031-2}
\lambda = \mathsf{\Lambda}(\sigma, \rho).   
\ee

Growth occurs by two mechanisms, one leading to an increase in $\ell_R$, the other to an increase in $\rho$.  Typically, the specimen involves ``load-carrying filaments'' that extend from one support to the other. Monomers from the surrounding monomer pool can be added to the tips of such filaments at the functionalized support at $x_0$.  This changes their reference length $\ell_R$.  There may also be ``free filaments'' in the specimen, for example with one end attached to the existing polymer network and the other free. Monomers can be added to the free ends of such filaments.  When a free filament grows, it may eventually touch one of the supports and turn into a load-carrying filament of the first type. Such growth changes the density $\rho$ of load-carrying filaments.

We assume that the rate of increase of the number of load-bearing filaments, $\dot\rho$, is governed by a kinetic law of the form
\be
\label{eq:20191031-5}
\dot \rho = \mathsf{R}(\rho),  
\ee
where the kinetic response function $\mathsf{R}$ is allowed to depend on $\rho$ but not  $\sigma$. 
Since  a ``free filament'' remains stress-free while it  grows by polymerization at its free tip, such growth is expected to be unaffected by stress and hence we have taken $\mathsf{R}$ to be independent of $\sigma$.  
As for surface growth due to polymerization, we assume that the speed of the moving boundary in reference space is related to the filament force by a  kinetic relation of the form
\be
\label{eq:20191031-4}
V = {\mathsf{V}}(f) = {\mathsf{V}}(\sigma, \rho).   
\ee
Since the kinetic response function $\mathsf{V}$ is a function of the filament force, it depends on both stress and filament density.   Similar to earlier work~\cite{mogiloster}, we write $\mathsf{V}$ as a function of the (average) filament force 
$f = \sigma/\rho$ to facilitate comparison with kinetic laws such as equation (\ref{eq:expokin}) that are given in the literature in terms of $f$ for single growing filaments \cite{howardbook}.

The 5 variables $\ell(t), \ell_R(t), \sigma(t), \lambda(t)$ and $\rho(t)$ are governed by the equations \eqref{eq-3xx}, \eqref{eq:20191031-2}, \eqref{eq:20191031-5} and \eqref{eq:20191031-4}, together  with \eqref{eq-2xx}, \eqref{eq-2yy}  and a loading condition such as the prescription of the stress history.



\section{Growth at constant filament density.} \label{sec-3}

In this section we model the test specimen as involving a fixed number of  nearly parallel filaments as in the experiments of Brangbour {\it et al.} \cite{brangbour}. The filament density therefore does not change and the constitutive relation of the specimen can be deduced from that of a single filament. Moreover, there is no cross-linking and filament bending is not important.

\subsection{Further details of the model.} \label{subsec-3-1}

\begin{figure}[h]
\centerline{\includegraphics[scale=0.35]{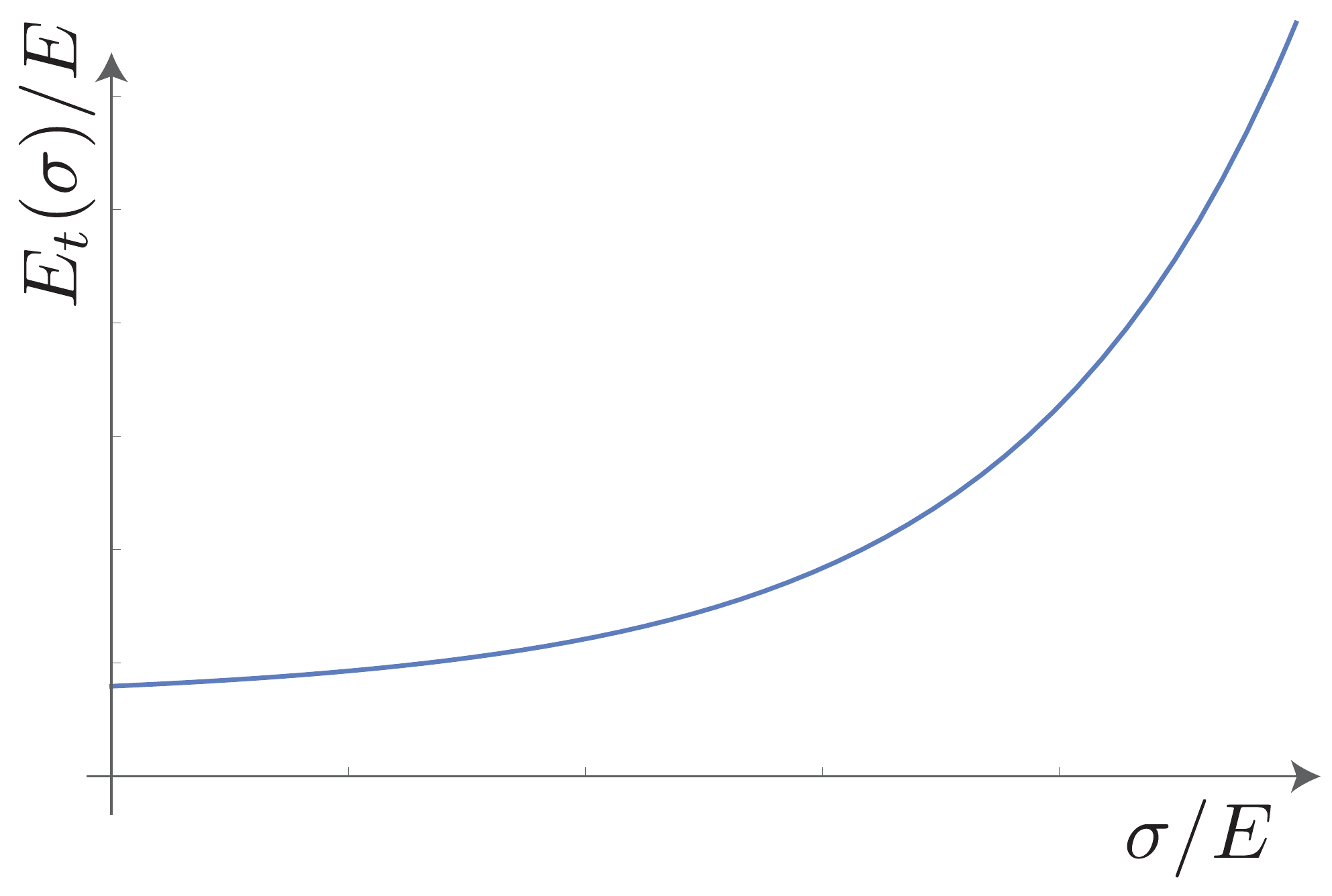}}
\caption{{Tangent modulus $E_t(\sigma)/E$ versus stress $\sigma/E$ according to  \eqref{eq:0117-31}.  The horizontal axis has been drawn on a logarithmic scale so as to facilitate comparison with Figure 3 of \cite{ChaudhuriParekh2007}.}}
\label{Fig-E-Sig.pdf}
\end{figure}

We now make the following choices for the constitutive response functions $\mathsf{\Lambda}, \mathsf{V}$ and $\mathsf R$.
The stress $\sigma$ in the network is a function of the stretch $\lambda$ that vanishes when $\lambda=1$ and becomes unbounded at extreme compression when $\lambda \to 0^+$. Moreover the material stiffens with increasing compressive stress. A simple model capturing this is 
\be
\label{eq:0117-21} 
\sigma = E(\lambda^{-1} - 1), \qquad \lambda = \mathsf{\Lambda}(\sigma, \rho) = \frac{1}{1 + \sigma/E} \qquad {\rm where} \quad E=E(\rho) \coloneqq \rho E_{0}
\ee
is the effective Young's modulus of the specimen. Note that the network elastic modulus is chosen to be (proportional to the number $N$ of filaments and therefore) linear in the filament density $\rho$ since the elasticity of the network in \cite{brangbour}
is of entropic origin such that $E_{0} \propto k_{B}T$.
The tangent modulus $E_t$ -- the slope of the stress-stretch curve --  is given as a function of stress by
\be
\label{eq:0117-31}
E_t(\sigma) \, {=} \,  - \left. \frac{d\sigma}{d\lambda}\right|_{\lambda = [1 + \sigma/E]^{-1}}\,  
=  E( 1 + \sigma/E)^2.
\ee
As shown in Figure \ref{Fig-E-Sig.pdf},  the graph of $E_t(\sigma)$ versus $\sigma$ according to \eqref{eq:0117-31} rises monotonically and is concave upwards which agrees with the trends observed experimentally, e.g. see Figure 3 of Chaudhuri {\it et al.} \cite{ChaudhuriParekh2007}.

In this section, the number $N$ of load-bearing filaments, and therefore the filament density $\rho$, is assumed to remain constant during each experiment. Thus
$$
\dot \rho = \mathsf{R}(\rho) = 0.
$$
Next, suppose that growth occurs at a constant rate $V_0$ provided the filament force $f$ is less that a certain critical value $f_{stall}$, or equivalently, when the stress is less than a certain value $\sigma_{stall}$.  Accordingly we take the kinetic law for surface growth \eqref{eq:20191031-4}  to be 
\be
\label{eq-0117-2}
V = \mathsf{V}(f) = \left\{
\ba{lll}
V_0 \qquad &{\rm for} \quad &f < f_{stall},\\[2ex]

0 \qquad &{\rm for} \quad &f > f_{stall},\\
\ea
\right.
\qquad \qquad
V =  \left\{
\ba{lll}
 V_0 \qquad &{\rm for} \quad &\sigma < \sigma_{stall},\\[2ex]

0 \qquad &{\rm for} \quad &\sigma > \sigma_{stall},\\
\ea
\right.
\ee
where $\sigma = \rho f, \sigma_{stall} = \rho f_{stall}$. This piecewise constant kinetic relation is depicted by the dotted line in Figure \ref{Fig-KineticLaws}. Such  kinetic relations arise in other 
mechanics settings where they have been 
associated with a notion of maximum dissipation, e.g. plasticity \cite{lubliner1984,rice1970}, phase transitions  \cite{rajkk1992} and kink band motion \cite{ogden2020}. They are referred to as {\it maximally dissipative kinetics}.


\subsection{Predictions of the model.} \label{subsec-3-2}

In this section we shall consider various predictions of the model described above.  Most of the results can be readily and conveniently described using nondimensional variables. However, we shall continue to use the (dimensional) variables introduced above since that allows us to make quantitative comparisons with the experimental measurements of Brangbour {\it et al.} \cite{brangbour}. A brief description of their experiments and a discussion of how their apparently different model relates to ours, is postponed to Section \ref{sec-brangmodel}. Details on how the various parameter values were chosen is described in Section S1 of the  Supplementary Material.

By \eqref{eq-3xx} and \eqref{eq:0117-21} the length $\ell$ of the specimen at any instant is related to the stress and unstressed reference length at that instant by
\be
\label{eq-0117-1}
\ell(t) = \frac{\ell_R(t)}{1 + \sigma(t)/E}.
\ee

\begin{figure}[h]
\centerline{\includegraphics[scale=0.25]{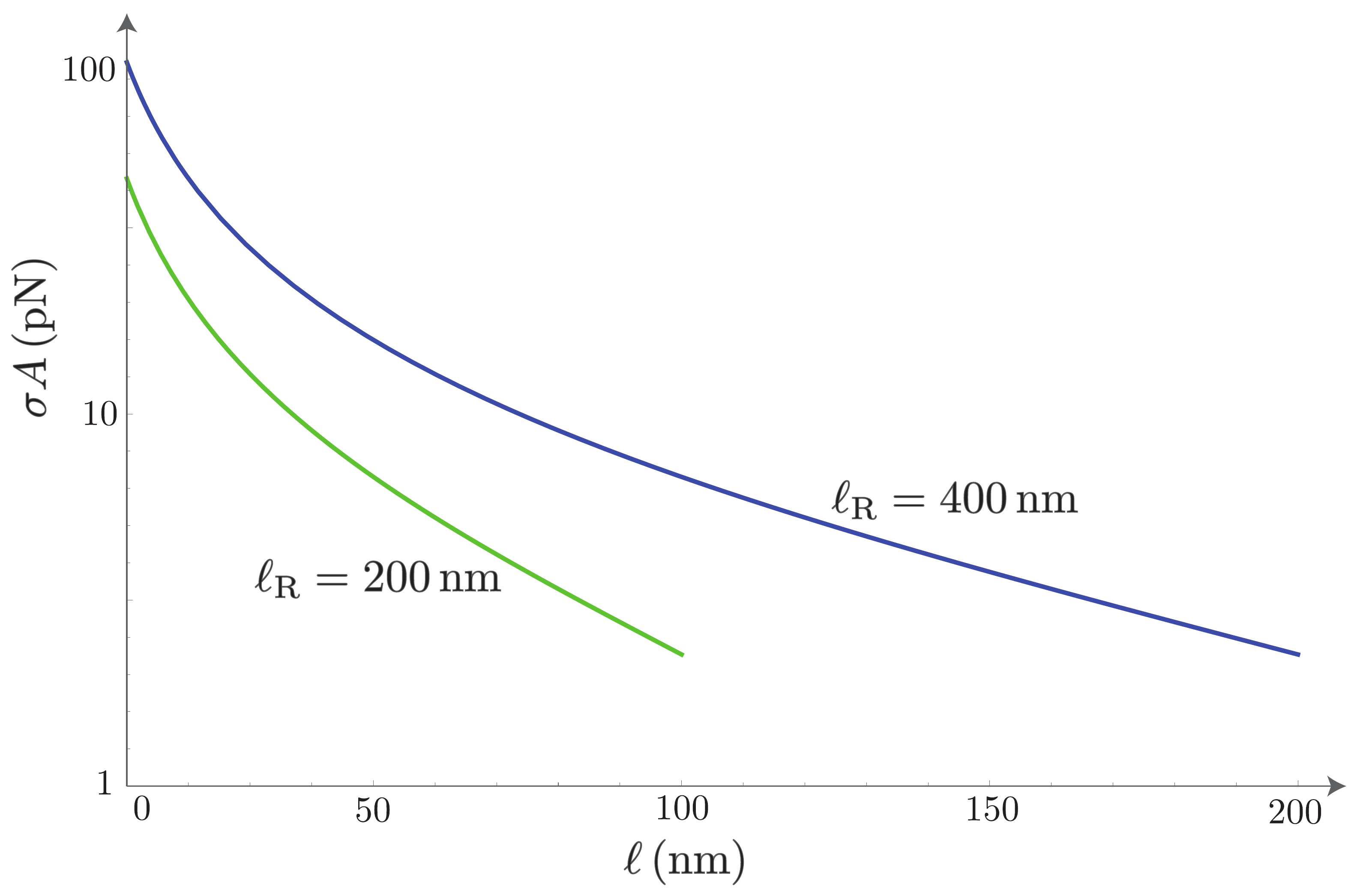}}
\caption{{Force $\sigma A$ versus length $\ell$ according to \eqref{eq-0117-1} when the loading-rate is much faster than the growth-rate allowing $\ell_R$ to be treated as constant. The figure has been drawn for $EA = 2.259 \, \si{pN}$ and two different values of $\ell_R$. The vertical axis is on a logarithmic scale. Compare with Figure 3 of \cite{brangbour}.}}
\label{Fig-ell-sig.pdf}
\end{figure}

In the {\it first} set of calculations, the filaments are initially permitted to grow freely under zero stress for some time interval $[0, t_0)$. During this stage, the unstressed reference length of the specimen is $\ell_R(t) =  V_0 t$ as required by \eqref{eq-0117-2},  \eqref{eq-2xx} and  \eqref{eq-2yy}. At time $t_0$ the force on the specimen is increased rapidly, and the force $\sigma A$ and specimen length $\ell$ are measured.  If the loading-rate is much faster than the rate of growth, the value of  $\ell_R$ can be assumed to remain constant at the value $\ell_R = V_0 t_0$.  Figure \ref{Fig-ell-sig.pdf} shows a plot of the force $\sigma A$ versus the specimen length $\ell$ in such an experiment as predicted by 
\eqref{eq-0117-1}  for two different fixed values of $\ell_R = V_0 t_0$. This figure may be compared with Figure 3 of Brangbour {\it et al.} \cite{brangbour}. 
%

\begin{figure}[h]
\centerline{\includegraphics[scale=0.25]{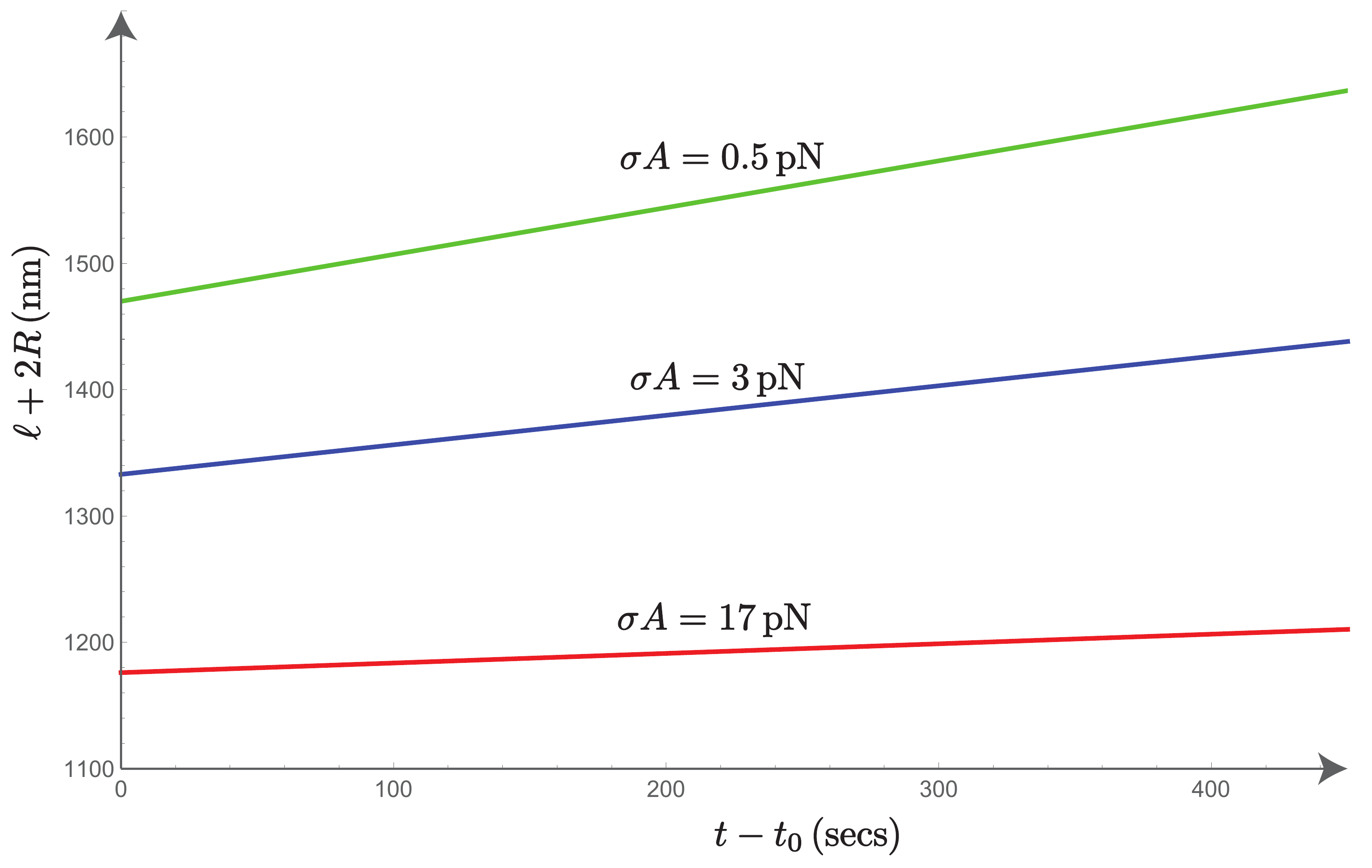}}
\caption{{The length of the specimen $\ell$ versus time $t$ according to \eqref{eq-0117-3} at two different values of the force $\sigma A$.  In order to compare this with Figure 1c of 
Brangbour {\it et al.} \cite{brangbour}, we have plotted $\ell + 2R$ on the vertical axis where $R=550\, \si{nm}$. The figure has been drawn using $V_0 = 0.42\, \si{nm/sec}, t_0 = 1000 \, \si{sec}$ and $EA=3.764\, \si{pN}$. The three lines correspond to $\sigma A =0.5\, \si{pN}$; $\sigma A =3\, \si{pN}$; and $\sigma A = 17 \, \si{pN}$.}}
\label{Fig-ell-t.pdf}
\end{figure}

In the {\it second} set of calculations,  the filaments are again permitted to grow freely under zero stress for some initial  time interval $[0, t_0)$ during which $\ell_R(t) = V_0 t$. At time $t_0$ a force is applied on the specimen and held constant. Then from
\eqref{eq-0117-2} and \eqref{eq-0117-1}, provided $\sigma <\sigma_{stall}$, 
\be
\label{eq-0117-3}
\ell(t) = \frac{V_0 t}{1 + \sigma/E}, \qquad t > t_0.
\ee
Figure \ref{Fig-ell-t.pdf} shows a plot of the specimen length $\ell(t)$ versus time $t$ for three different fixed values of the force $\sigma A$. Observe that the slope decreases as the force increases. This figure may be compared with Figure 1c of Brangbour {\it et al.} \cite{brangbour}.
%

\begin{figure}[h]
\centerline{\includegraphics[scale=0.25]{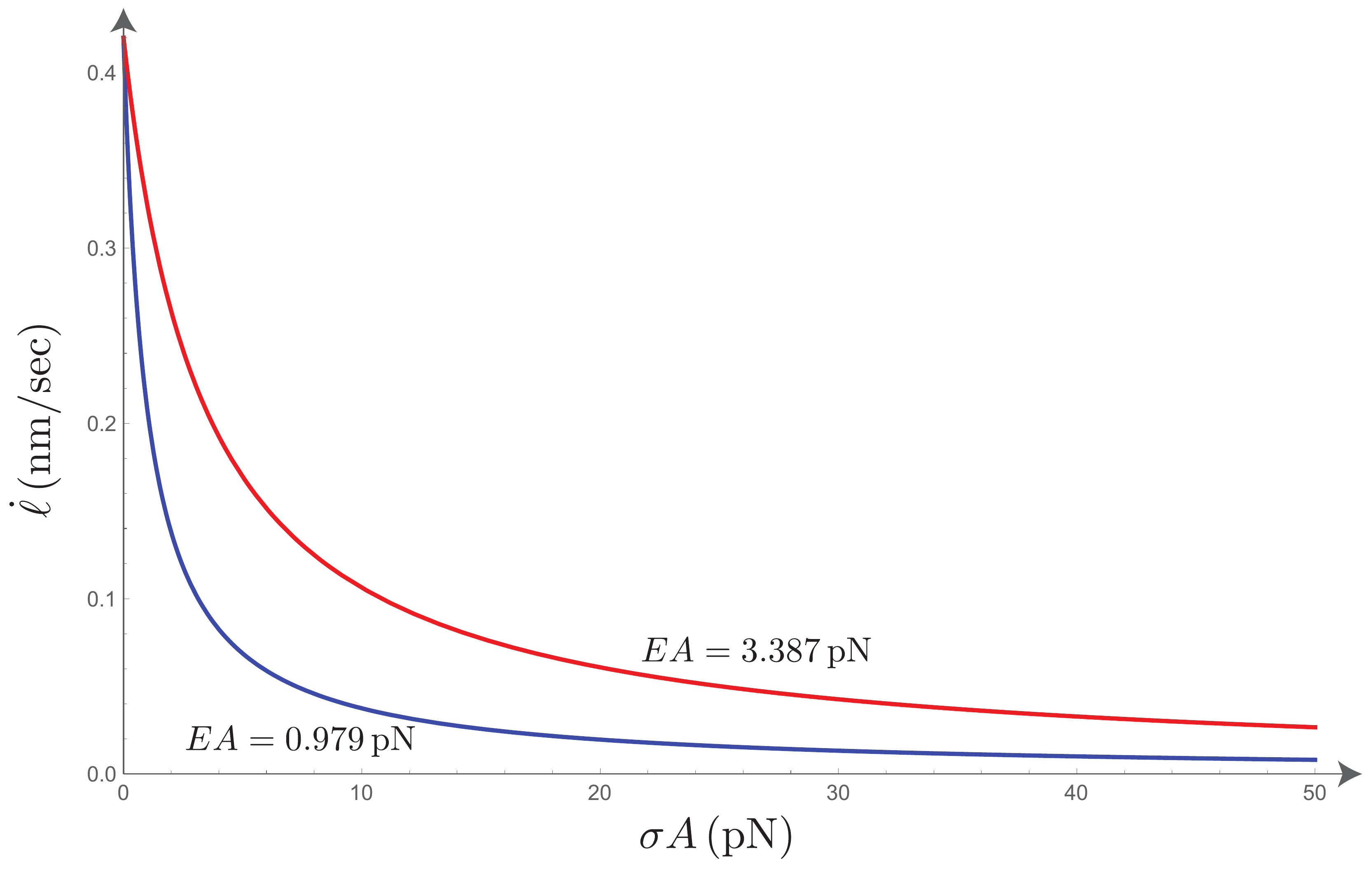}}
\caption{{Elongation-rate $\dot \ell$ versus force $\sigma A$ at two different values of $E A$ according to  \eqref{eq-0117-4}. The figure has been drawn for $V_0 = 0.42 \, \si{nm/sec}$. Compare with Figure 5 of \cite{brangbour}.}}
\label{Fig-elldot-sig2.pdf}
\end{figure}

The elongation rate-force response\footnote{frequently referred to as the force-velocity relation in the literature,} plays a central role in the study of actin filament growth.  Thus consider a {\it third} set of calculations in which the force is again held constant but now examine the elongation-rate $\dot\ell$ as a function of force $\sigma A$ and in particular on how it depends on $EA$. Differentiating \eqref{eq-0117-3} with respect to $t$ at constant $\sigma$ gives
\be
\label{eq-0117-4}
\dot \ell = \frac{V_0}{1 + \sigma A/(EA)},
\ee
and Figure \ref{Fig-elldot-sig2.pdf} shows a plot of the elongation-rate $\dot \ell$ versus the force $\sigma A$ according to \eqref{eq-0117-4}. The two curves correspond to  $EA = 3.387 \, \si{pN}$ and $EA = 0.979 \, \si{pN}$. This figure may be compared with Figure 5 of Brangbour {\it et al.} \cite{brangbour}.
%


{\it Fourth,} suppose that the specimen is initially growing under some constant stress $\sigma_1$ and that at some instant $t_1$ the stress is suddenly increased to a much larger value $\sigma_2$ (less than the stall stress). The stress is kept  constant at this higher value until time $t_2$ at which instant it is suddenly decreased back to its original value $\sigma_1$ and held constant at that value from then on.  This is described by  the loading history
\be
\label{eq-0118-1}
\sigma(t) = \left\{
\ba{lll}
\sigma_1 \qquad &{\rm for} \quad &0 < t < t_1 ,\\[2ex]

\sigma_2 \qquad &{\rm for} \quad &t_1 < t < t_2,\\[2ex]

\sigma_1 \qquad &{\rm for} \quad &t > t_2,\\
\ea
\right.  \qquad {\rm where} \quad 0 < \sigma_1 \, < \,  \sigma_2 \ (< \sigma_{stall}).
\ee
The corresponding response is found by solving  \eqref{eq-0117-1}, \eqref{eq-0117-2}, \eqref{eq-2xx} and \eqref{eq-2yy}:
\be
\label{eq:0118-10}
\ell(t) = \frac{\ell_R(t)}{1 + \sigma(t)/E}, \qquad \dot\ell_R(t) = V_0.
\ee
In solving \eqref{eq:0118-10} it is important to keep in mind that the referential length of the filament $\ell_R(t)$ changes {\it only} due to growth. Thus at an instant at which the stress changes discontinuously, the length $\ell(t)$ of the filament will also change discontinuously due to the jump in $\sigma$ but the referential length will remain continuous since a finite segment of new material cannot appear in infinitesimal time.  Thus from \eqref{eq-0118-1} and \eqref{eq:0118-10}
\be
\label{eq-0118-2}
\ell(t) = \left\{
\ba{lll}
\frac{V_0 t}{1 + \sigma_1/E} \qquad &{\rm for} \quad &t_0 < t < t_1 ,\\[2ex]

\frac{V_0 t}{1 + \sigma_2/E} \qquad &{\rm for} \quad &t_1 < t < t_2,\\[2ex]

\frac{V_0 t}{1 + \sigma_1/E} \qquad &{\rm for} \quad &t > t_2,\\
\ea
\right. 
\qquad
\dot\ell(t) = \left\{
\ba{lll}
\frac{V_0}{1 + \sigma_1/E} \qquad &{\rm for} \quad &t_0 < t < t_1 ,\\[2ex]

\frac{V_0}{1 + \sigma_2/E} \qquad &{\rm for} \quad &t_1 < t < t_2,\\[2ex]

\begin{figure}[h]
\centerline{\includegraphics[scale=0.3]{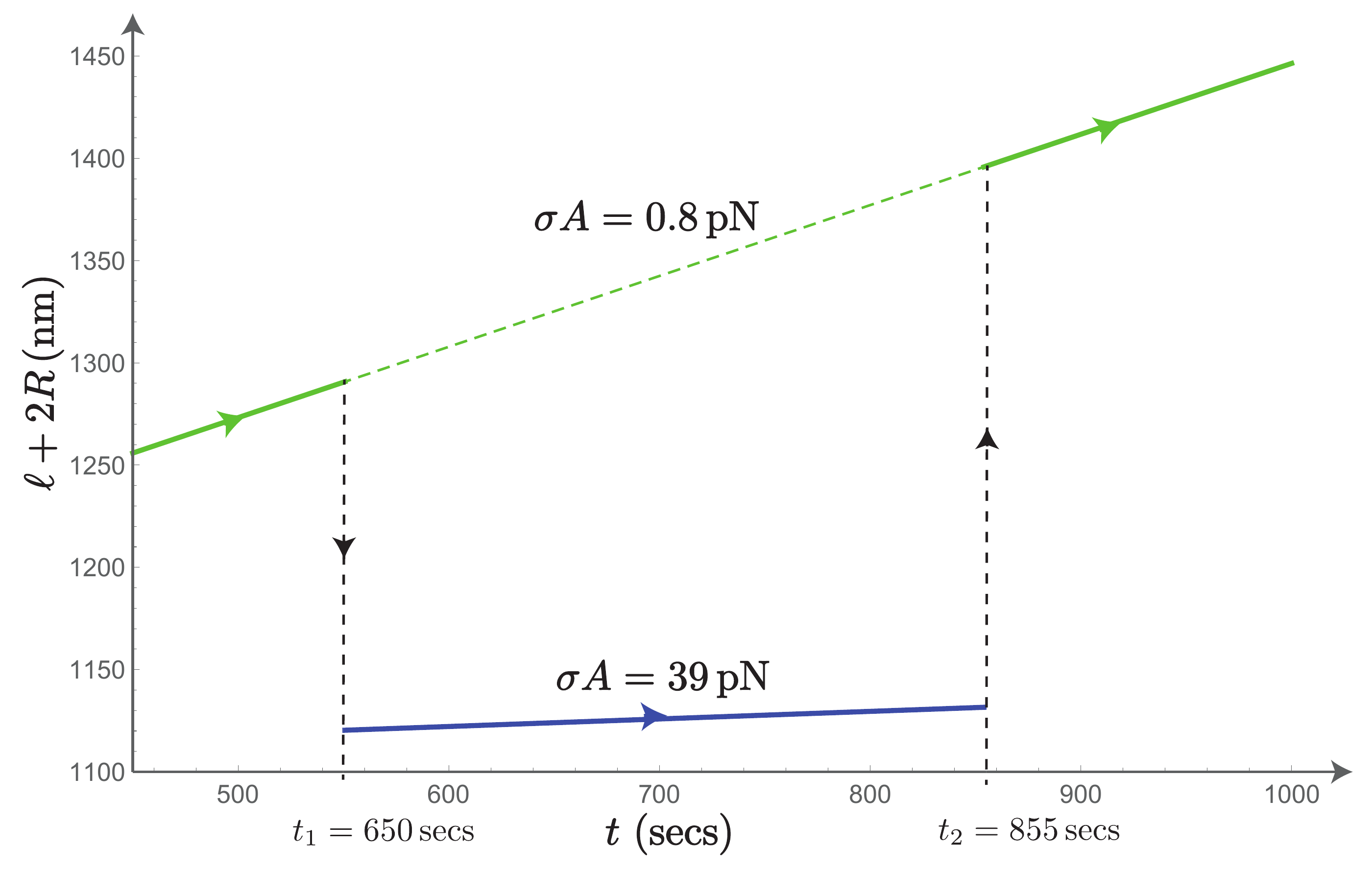}}
\caption{{The length of the specimen $\ell$ versus time $t$ according to \eqref{eq-0118-2}$_1$ with $\sigma_1 A = 0.8\, \si{pN}$ and $\sigma_2 A = 39 \, \si{pN}$.   In order to compare this with Figure 1c of Brangbour {\it et al.} \cite{brangbour}, we have plotted $\ell + 2R$ on the vertical axis where $R=550\, \si{nm}$. The figure is drawn for $V_0 = 0.42\, \si{nm/sec}$ and $EA =  3.764 \, \si{pN}$.  See also Figure \ref{Fig-ell-t.pdf}. }}
\label{Fig-ell-t-jumps.pdf}
\end{figure}

\frac{V_0}{1 + \sigma_1/E} \qquad &{\rm for} \quad &t > t_2.\\
\ea
\right. 
\ee
In Figure \ref{Fig-ell-t.pdf} we observed that there is effectively a one-parameter family of straight lines on the $t, \ell$-plane, the force $\sigma A=F$ being the parameter.  During the loading \eqref{eq-0118-1}, the point $(t, \ell(t))$ first moves on the straight line corresponding to $\sigma A = F_1$ in Figure \ref{Fig-ell-t-jumps.pdf}. At time $t_1$ it jumps down to the  straight line associated with $\sigma A = F_2$ and traverses that line for $t_1 < t < t_2$, and finally at time $t_2$ it jumps back up to the straight line $\sigma A = F_1$ where $F_i$ is the force. This figure may be compared with Figure 2 of Brangbour {\it et al.} \cite{brangbour}.
%


\subsection{The model of Brangbour {\it et al.} \cite{brangbour}.} \label{sec-brangmodel}

Brangbour {\it et al.} \cite{brangbour} devised a novel method for measuring the force-velocity relation for a system involving a relatively small number of growing actin filaments. They used a suspension containing (actin monomers as well as) magnetic colloidal particles that when subjected to a magnetic field assembled into a linear chain.  The surfaces of the colloidal particles were functionalized so that a certain controlled number of actin filaments grew radially from the surface of each particle. The interaction between the filaments on two adjacent particles caused the particles to move apart against the magnetically applied force. In this way the authors measured the force, the separation between particles and their velocity. Several of their results were described above.

 \begin{figure}[h]
\centerline{\includegraphics[scale=0.30]{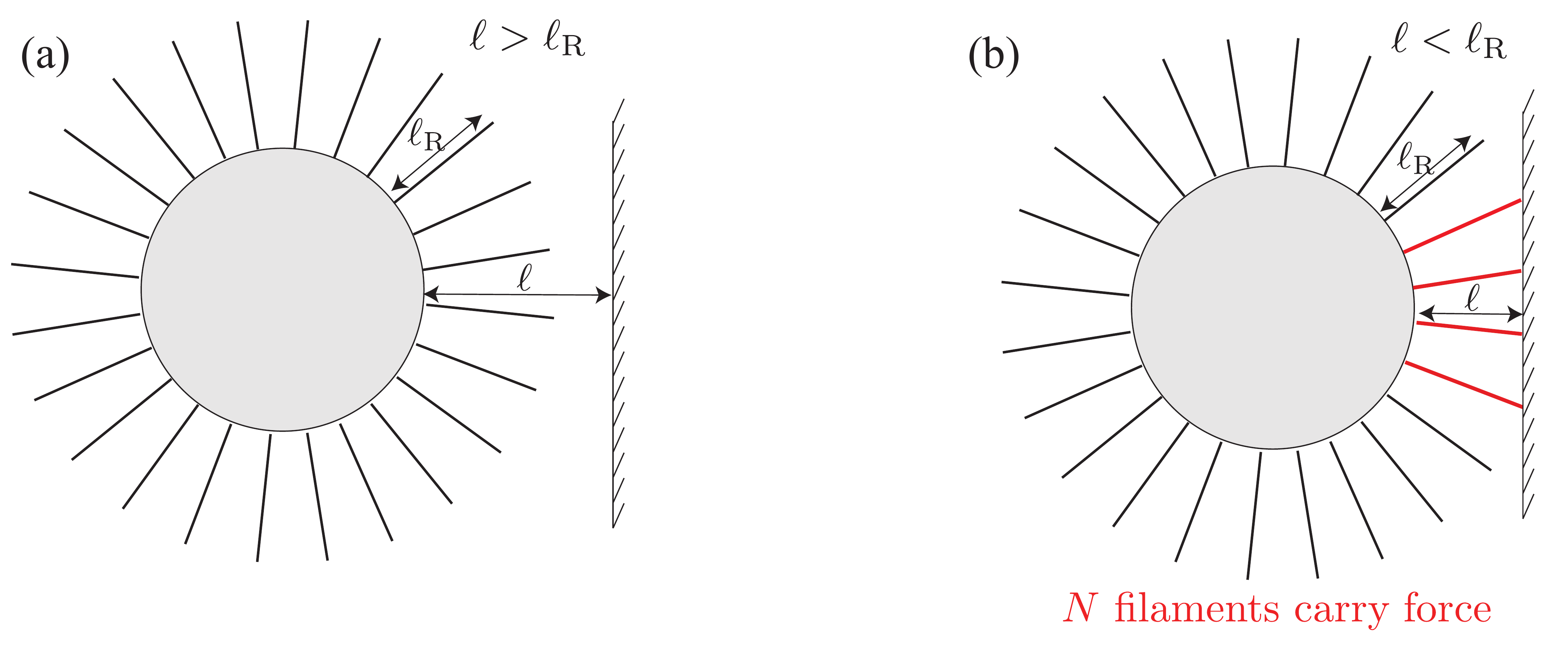}}
\caption{{(a) Case $\ell > \ell_R$:  The filaments are not in contact with the wall and so remain stress-free. (b) Case $\ell < \ell_R$: each of $N$ filaments carry a force $f$. Both $N$ and $f$ increase as $\ell$ decreases (at fixed $\ell_R$). }}
\label{Fig-Brangbour-Expt2}
\end{figure}

Brangbour {\it et al.} \cite{brangbour} arrive at (what is effectively) equation \eqref{eq:0117-21} using a more accurate version of the following simplified micro-mechanical argument:  suppose that actin filaments of stress-free length $\ell_R$ are attached to a rigid spherical particle of radius $R$ that is in the vicinity of a wall. The wall represents the mid-plane between a pair of adjacent particles.  When  the distance $\ell$ from the particle surface to the wall is greater than $\ell_R$ as in Figure \ref{Fig-Brangbour-Expt2}(a), there is no contact between the filaments and the wall and so the filaments carry no force.  On the other hand if the distance to the wall is less than $\ell_R$ as in Figure \ref{Fig-Brangbour-Expt2}(b), some filaments will interact with the wall and carry force.  The (compressive) force $f$ in a filament increases as the distance $\ell$ to the wall decreases. Suppose the force is given by the classical entropic model
$$
f = {c_1}/{\ell} \qquad {\rm for} \quad \ell < \ell_R,
$$
where the parameter $c_1$ is related to the elastic modulus of a filament.  Moreover, as the distance between the particle and wall decreases, the number, $N$, of filaments interacting with the wall, and therefore carrying force, increases  from the value $N=0$ at $\ell = \ell_R$.  Assume that the number of force carrying filaments is given by the linear relation
$$
N = c_2(\ell_R - \ell) \qquad \mbox{for} \quad \ell < \ell_R,
$$
where the parameter $c_2$ is related to the density of filaments on the particle surface. The total force between the particle and wall, $fN$, when calculated using the two preceding equations, leads to precisely a  constitutive relation of the form \eqref{eq:0117-21} with
$$
EA = c_1 c_2.
$$
{\it It is worth emphasizing that when the constitutive relation \eqref{eq:0117-21} is derived in this way, it accounts for both the stress-stretch behavior of a filament and the changing number of load carrying filaments.} 

According to the entropic model $c_1 = k_BT$ where $k_B$ is the Boltzmann constant and $T$ is the absolute temperature, and the linearization of the geometric relation derived by  Brangbour {\it et al.} \cite{brangbour}  gives $c_2 = N_{GS}/(4R)$ where $N_{GS}$ is the fixed number of filaments on a colloidal particle. Therefore one finds
$$
EA = c \, k_BT \, \frac{N_{GS}}{4R}
$$
where $c$ is a factor they introduce to better fit the data.  Observe from this that the effective modulus $EA$ can be varied by changing the density of filaments on the particle surface. 

Our equation \eqref{eq-0117-1} is {\it identical} to equation $(4)_2$  in Brangbour {\it et al.}
\cite{brangbour} provided we identify their variables $X, V_0 t$ and $ ck_BTN_{GS}/(4R)$ with our $\ell, \ell_R$ and $E$ respectively.



\section{Growth of a network of actin filaments.} \label{sec-4}

\begin{figure}[h]
\centerline{\includegraphics[scale=0.75]{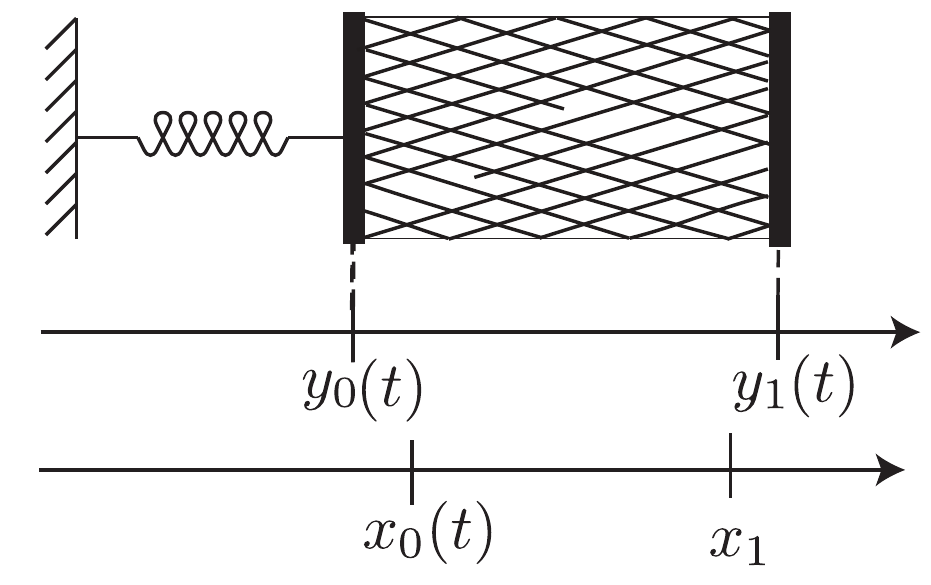}}

\caption{{Schematic figure of test specimen depicting cross-linking, load-bearing filaments and free filaments. The specimen is identified with the interval $[y_0(t),y_1(t)]$ in physical space. It is attached to an AFM-spring at its left-hand end $y_0(t)$ and to a (movable) support at its right-hand end $y_1(t)$.   In reference space, the specimen is identified with the interval $[x_0(t),x_1]$.   Surface growth occurs at the left end of the specimen causing the boundary $x=x_0(t)$ to move leftward.}}

\label{Fig-TestSpecimen.pdf}
\end{figure}

We continue to consider the test specimen described in Section \ref{sec-2} but now with its left end attached to an AFM cantilever and its right end to a (movable) support as depicted schematically in Figure \ref{Fig-TestSpecimen.pdf}.  The AFM acts like a linear spring whose force-elongation relation is
\be
\label{eq-20200131-11}
\sigma A = - k_c(y_0 - Y_0).  
\ee
Here $\sigma(t)$ is the average (compressive) stress in the specimen, the cross-sectional area of the specimen over which the filaments are distributed is  $A$, $Y_0$ is the position of the AFM cantilever when it is undeflected, $y_0(t)$ is the position of the left-hand end of the specimen at time $t$ and $k_c$ is the (usual) spring stiffness in units of force/displacement. It is more convenient to set $k = k_c/A$ and write \eqref{eq-20200131-11} as
\be
\label{eq-4xx}
\sigma  = - k(y_0 - Y_0). 
\ee


\subsection{Constitutive response functions.} \label{sec:20191220-1}


\subsubsection{Stress-stretch-filament density relation.}   \label{sec-20200210-1} 

First consider the constitutive function $\mathsf{\Lambda}$  describing the relation between the stress $\sigma$, stretch $\lambda$ and filament density $\rho$. Suppose that the force $f$ in a filament is related to its stretch  by $f = E_fA_f(\lambda^{-1} - 1)$ where $E_f$ is its Young's modulus and $A_f$ its cross-sectional area.  
This, together with \eqref{eq-0505-1}, gives the constitutive relation for the specimen to be $\sigma = f\rho = E(\lambda^{-1} - 1)$ where $E= \rho E_fA_f$.

The filaments in the experiments of Brangbour {\it et al.} \cite{brangbour} are relatively short and do not form a network.  In contrast, the actin filaments in the experiments of Parekh {\it et al.}  \cite{ParekhChaudhuri2005} form a cross-linked network where filament bending becomes important. To account for an analogous phenomenon in foams, Gibson and Ashby  proposed the modification
$E= \rho^2A_f^2 E_f$
to the effective Young's modulus, see \cite{gibsonAshbybook}.  We adopt their model and write the stress-stretch-filament density relation in the form  $\lambda = \mathsf{\Lambda}(\sigma, \rho)$ where 
 \be
  \label{eq:20200125-1}
\mathsf{\Lambda}(\sigma, \rho) = \frac{1}{1 + \sigma/E},  \qquad E = E(\rho) = \rho^2A_f^2E_f. 
 \ee
For a three-dimensional continuum model of actin networks see for example \cite{boyce2008}.

Keeping in mind that we are taking compressive stress to be positive, we are concerned with $\sigma \geq 0$ whence $\lambda \leq 1$.  Since the total cross-sectional area taken up by the filaments, $NA_f$, cannot exceed the cross-sectional area $A$ over which the filaments are distributed, it is necessary that
    \be
   \label{eq-16xx}
   0 \leq \rho \leq 1/ A_f.  
   \ee

  
\subsubsection{Kinetic law for filament density.}   \label{sec-20200210-2}

Next consider the kinetic relation $\dot \rho = \mathsf{R}(\rho)$ governing the filament density. As the number of filaments increases,  the number of monomers in the surrounding monomer pool decreases, and so the rate at which new filaments develop is expected to decrease, i.e. we anticipate $\dot \rho$ to be a decreasing function of $\rho$. Moreover, since the total number of monomers in the system is finite, as well as because of \eqref{eq-16xx}, $\rho$ cannot increase indefinitely.
Therefore for the kinetic law $\dot \rho= \mathsf{R}(\rho)$ we take
\be
\label{eq:20191219-1}
\tau_\rho\, \dot \rho = \rho_\infty - \rho, \qquad \tau_\rho >0, \quad \rho_\infty >0,  
\ee
where $\tau_\rho$ and $\rho_\infty$ are constant parameters.  The linear dependence of $\dot{\rho}$ on $\rho$ is similar to that in \cite{mogiloster}. 
Since one can solve \eqref{eq:20191219-1} explicitly, the response predicted by this kinetic relation is 
\be
\label{eq:20191219-3}
\rho(t) = \rho_{\infty} + (\rho_0 - \rho_\infty)\, e^{-(t-t_0)/\tau_\rho},  
\ee
where $\rho_0 = \rho(t_0) < \rho_\infty$ is the filament density at some particular instant $t_0$. Note that $\rho(t)$ increases monotonically and $\rho(t) \to \rho_\infty$ as $t \to \infty$.

Since $\rho(t) < \rho_\infty$, this, together with  $E = \rho^2 A_f^2E_f$ and $\rho(t) \to \rho_\infty$ as $t \to \infty$,  tell us that the effective Young's modulus obeys
$$
E(t) < E_\infty \quad {\rm and} \quad E(t) \to E_\infty \quad {\rm as} \quad t \to \infty \qquad {\rm where} \  E_\infty \coloneqq \rho_\infty^2 A_f^2 E_f.
$$
Thus $E_\infty$ is the Young's modulus when the system reaches steady state.  It will be convenient to write
 \eqref{eq:20200125-1}$_2$ in terms of $E_\infty$ as
\be
\label{eq-20200209-2}
E = E_\infty \, \rho^2/\rho^2_\infty.
\ee


\subsubsection{Kinetic law for surface growth.}  \label{sec-20200201-1}

We now present two models for the kinetic relation $V = \mathsf{V}(f) = \mathsf{V}(\sigma, \rho)$ describing surface growth at the left-hand boundary of the specimen where $V = - \dot x_0$.  
According to the literature, e.g. \cite{brangbour,ParekhChaudhuri2005},   growth is expected to stall at some critical value $f_{stall}$ of the filament force. 
Therefore we take $\mathsf{V}(f)$ to be a monotonically decreasing function of $f$ with  $\mathsf{V}(f) \to 0$ as $f \to f_{stall}$.

\begin{figure}[h]
\centerline{\includegraphics[scale=0.30]{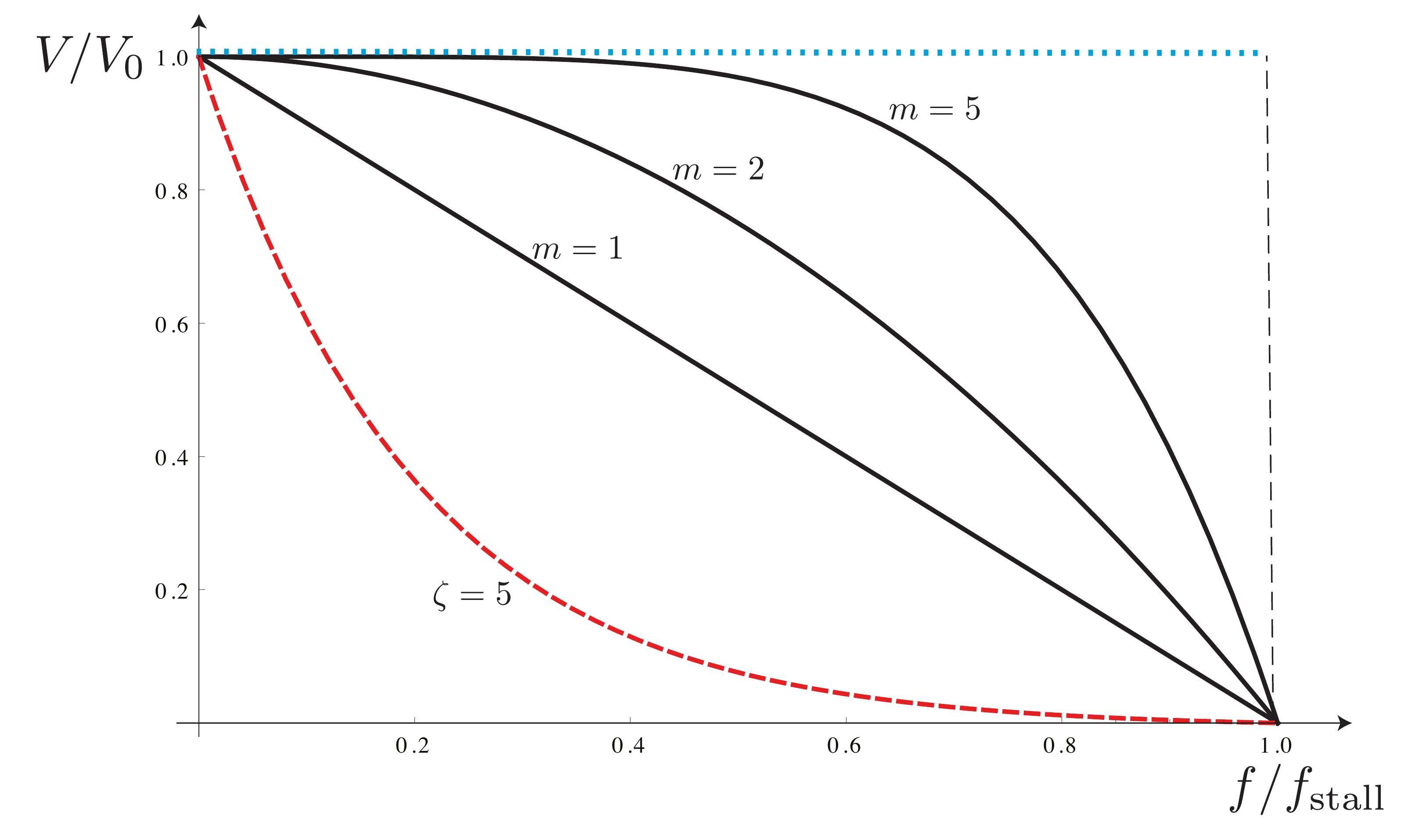}}
\caption{{Growth speed $V/V_0$ versus filament force $f/f_{stall}$ for surface growth. Dotted blue: maximum dissipation kinetic law \eqref{eq-0117-2}$_1$ from Section \ref{sec-3}. Dashed red: exponential kinetic law \eqref{eq:20191111-3} with $\zeta= 5$.  Solid black: power-law kinetics according to \eqref{eq:20191206-3} with $m=1, 2$ and $5$. }}
\label{Fig-KineticLaws}
\end{figure}

We first present the kinetic law (of Arrhenius form) that is exponential in the filament force $f$:
\be
\label{eq:20191111-3}
V = V_0 \frac{ \left[ e^{-\zeta \, f/f_{stall}} - e^{- \zeta}\right] }{ \left[1 - e^{- \zeta}\right]} \qquad {\rm for} \quad 0 \leq f \leq f_{stall}, 
\ee
where $f_{stall}, V_0$ and $\zeta$ are constant parameters with
$$
V_0>0, \qquad f_{stall} >0, \qquad \zeta \neq 0. 
$$ 
For reasons that we will explain below, we shall {\it not} use this  kinetic relation in our calculations but note it here because of its frequent use in this field. Howard \cite{howardbook} gives an expression for the parameter $\zeta$ in terms of the temperature $T$, the Boltzmann constant $k_B$ and the length of a stress free monomer $a$:
\be
\label{eq-howard}
\zeta = \frac{a f_{stall}}{k_BT},\
\ee
indicating that the parameter $\zeta$ should also be positive. 
Keeping in mind that the filament force $f$ is positive in compression, \eqref{eq:20191111-3} says that growth occurs ($V >0$) for compressive forces in the range $0 \leq f < f_{stall}$ and that growth stalls when $f \to f_{stall}$.  The dashed red curve in Figure \ref{Fig-KineticLaws} shows the variation of $V$ with $f$ according to \eqref{eq:20191111-3} for $\zeta = 5$. When $\zeta \to 0$ this curve approaches the straight line $V/V_0 =  1 - f/f_{stall}$.

Next, recall the maximum dissipation kinetic law for growth \eqref{eq-0117-2}$_1$ used in Section \ref{sec-3}: $V = V_0$ for $0 \leq f < f_{stall}$, $V = 0$ for $f > f_{stall}$. This is shown dotted in Figure \ref{Fig-KineticLaws}. In this section we adopt the following regularized (smoothed out) version of this kinetic law:
\be
\label{eq:20191206-3}
V = \mathsf{V}(f) = V_0 \left[ 1- \left(\frac{f}{f_{stall}}\right)^{m}  \right] \qquad {\rm for} \quad 0 \leq f \leq f_{stall}, 
\ee
where $f_{stall}, V_0$ and $m$ are constant parameters such that 
$$
V_0>0, \qquad  f_{stall} >0, \qquad m > 0. 
$$ 
For $m=1$ this model is linear in the filament force and for $m \to \infty$ it approaches the maximum dissipation kinetic law used in Section \ref{sec-3}. 
The solid curves in Figure \ref{Fig-KineticLaws} show the variation of $\dot\ell_R$ with $f$ according to \eqref{eq:20191206-3} for some different values of $m$.

The experimental observations in \cite{ParekhChaudhuri2005} indicate that the process of growth under spring loading involves an intermediate stage where the elongation-rate is almost independent of the stress, and therefore a plot of $\dot \ell$ versus $\sigma$ involves a more-or-less horizontal segment prior to stall. This is ideally captured by the maximum dissipation kinetic relation (which, as we saw in Section \ref{sec-3}, also did very well in modeling the experiments of Brangbour {\it et al.} \cite{brangbour}). Because the approach to stall is gradual and not sudden, a regularized version of that kinetic relation is desirable such as the power-law model \eqref{eq:20191206-3} with a moderate value of $m$. As can be seen from Figure \ref{Fig-KineticLaws}, the  exponential model \eqref{eq:20191111-3} does not capture this behavior.  The exponential kinetic relation {\it does} approach the maximum dissipation kinetic law when $\zeta \to - \infty$, though negative values of $\zeta$ appear not to be reasonable, e.g. Howard's model  \eqref{eq-howard}.  

Instead of the growth speed parameter $V_0$ it will sometimes  be more convenient to use the time-scale for growth defined by
\be
 \label{eq:20200131-25xz}
\tau_R \, \coloneqq \, {\ell_0}/{V_0},
\ee
where $\ell_0$ is the distance between the AFM cantilever and the other support when the AFM is undeflected.  When growth commences, the filaments attached to the AFM do not extend all the way to the other support. Therefore they initially grow under zero stress. Their length when they first touch the other support is $\ell_0$.

The implications of the dissipation inequality on the kinetic law for growth are discussed in the Appendix and Section S4 of the  Supplementary Material.


\subsubsection{Model parameters. Parameter values.}\label{subsubsec-parametervalues}

The constitutive models described in Sections  \ref{sec-20200210-1},  \ref{sec-20200210-2} and  \ref{sec-20200201-1} involve the following parameters: the stress-stretch-filament density relation \eqref{eq:20200125-1}$_1$, \eqref{eq-20200209-2} involves the Young's modulus $E_{\infty}$ and the maximum filament density $\rho_{\infty}$.  The kinetic law \eqref{eq:20191219-1} for the nucleation of new filaments involves $\rho_{\infty}$ and the time-scale $\tau_{\rho}$. The kinetic law for surface growth \eqref{eq:20191206-3},  \eqref{eq:20200131-25xz} involves the time scale $\tau_R$, the stall force $f_{stall}$ and the exponent $m$. Since stall occurs when $f=f_{stall}$ and $\rho=\rho_\infty$ it is convenient to define
$$
\sigma_{stall} \coloneqq \rho_\infty f_{stall}.
$$
In addition, the constitutive relation of the AFM spring involves its stiffness $k$. It is useful to define an associated stress $\sigma_{0}$ by 
\be
\label{eq-0501-sig0}
\sigma_{0} \coloneqq k \ell_0, 
\ee
where $k$ is the stiffness of the AFM spring in units of stress/displacement; see \eqref{eq-4xx}.

While it is natural to work with nondimensional variables and parameters, we shall not do so here since we want to make quantitative comparisons with the experimental results of Parekh {\it et al.}~\cite{ParekhChaudhuri2005}.   

How we arrived at the following specific values of the various parameters, including the sources of the data, is described in Section S2 of the  Supplemental Material.  Here we simply record the values we shall use:
\be
\label{eq:20200208-45}
\ba{cll}
\ell_0 = 3 \, \si{\mu m}, \quad  \sigma_{stall} = 0.77  \, \si{nN/\mu m^2}, \quad  \tau_\rho = 40  \, \si{min}, \quad  m=5,\\[2ex]

\sigma_0 = 0.2362  \, \si{nN/\mu m^2} \ (\mbox{Figures 2 and 3a of \cite{ParekhChaudhuri2005}}), \\[2ex]

\sigma_0 = 0.1575  \, \si{nN/\mu m^2} \ (\mbox{Figure 3b of \cite{ParekhChaudhuri2005}}). \\
\ea
\ee
The value of $\tau_\rho$ was chosen arbitrarily, while that of $m$ 
was chosen to ensure that the kinetic law $V = \mathsf{V}(f)$ was reasonably close to the maximum dissipation kinetic law as depicted in Figure \ref{Fig-KineticLaws}. This was necessary in order to get qualitative agreement between the theoretical predictions and the experiments. Two different AFM springs were used in \cite{ParekhChaudhuri2005} leading to the two values of $\sigma_0 = k \ell_0$ above. The value of $\rho_\infty$ turns out not to be needed but its value is of the order of $80$ filaments per $\si{\mu m^{2}}$, significantly smaller than the maximum filament density $1/A_f = 53,000 \, \si{\mu m^{-2}}$.

The two main sets of experiments carried out by Parekh {\it et al.}  \cite{ParekhChaudhuri2005} pertain to their Figures 2 and 3. 
It can be seen from those figures that the elongation-rate for growth under spring-loading is $\dot \ell = 72 \, \si{nm/min}$ in Figure 2 and $\dot \ell = 129 \, \si{nm/min}$ in Figure 3. Therefore the conditions under which the two sets of experiments were carried out had to be different, and so it is not unreasonable for the values of certain parameters in the model to also be different.  The elongation-rate $\dot \ell$ depends sensitively on the growth-speed $V_0$ which in turn depends on the monomer concentration in the surrounding solvent.  We assume that the different values of $\dot \ell$ observed is likely due to different monomer concentrations and so  take different values for $V_0$ (equivalently $\tau_R = \ell_0/V_0$) depending on which experiment we are modeling. Furthermore,
it can be readily shown from  \eqref{eq:8p} and \eqref{eq:20200108-3} below that the elongation-rate at the initial instant, $\dot \ell(0)$, depends sensitively on (the time scale for growth, $\tau_R$, and) the Young's modulus at steady-state, $E_\infty$. Therefore we also take the value of $E_\infty$ to be different in the two analyses.
When concerned with the experiments focused on growth under the action of the AFM spring (Figure 2 of \cite{ParekhChaudhuri2005}), we shall take
\be
\label{eq:20200208-46}
E_\infty = 3.7 \, \si{nN/\mu m^2}, \qquad \tau_R = 34  \, \si{min},
\ee
whereas when modeling the experiments focused on stress jumps (Figure 3 of \cite{ParekhChaudhuri2005}), we take
\be
\label{eq:20200208-47}
E_\infty = 0.7 \, \si{nN/\mu m^2},  \qquad \tau_R = 10  \, \si{min} \ \mbox{(Figure 3a)}, \qquad \tau_R = 13  \, \si{min} \ \mbox{(Figure 3b)}.
\ee
As can be seen from the  Supplementary Material, the values  in \eqref{eq:20200208-46} and \eqref{eq:20200208-47} are within the range of experimentally measured data. 


 \subsection{Growth under the action of the AFM spring.} \label{subsec-spring}

First consider the problem where the specimen grows under spring-loading.  Our main interest is in calculating the force,  specimen length and elongation-rate as functions of time, and then looking at a plot of elongation-rate versus force.

 In these calculations the  position $y_1$ of the support on the right-hand side is held fixed and the velocity $\dot y_0$ of the AFM cantilever  is measured. 
Thus the unstressed length of the specimen at the initial instant is
$ \ell_0  \coloneqq  y_1 - Y_0$
which represents the distance between the support and AFM cantilever when the AFM is not deflected.  It then follows because $\ell = y_1 - y_0$ that $\ell_0 - \ell = y_0 - Y_0$, and so the spring loading equation $\sigma  = -k(y_0 - Y_0)$ can be written as $ \sigma  = k(\ell - \ell_0)$:
$$
\ell = \ell_0 + \sigma/ k. 
$$

 The system of 4 equations to be solved to find $\ell(t), \ell_R(t), \sigma(t), \rho(t)$ are
 \be
   \label{eq:1p}
 \ell = \mathsf{\Lambda}(\sigma, \rho) \ell_R,\qquad  
 \dot \ell_R = \mathsf{V}(\sigma, \rho)  , \qquad
  \dot \rho = \mathsf{R}(\rho) , \qquad
\ell  = \ell_0 + \sigma/k ,
 \ee
having used $\dot \ell_R = - \dot x_0 = V$.  From \eqref{eq:1p}$_4$, 
 \be
  \label{eq:4p}
 \dot \sigma  = k \dot \ell.  
 \ee
Differentiating \eqref{eq:1p}$_1$ with respect to time and using \eqref{eq:4p} leads to
 \be
   \label{eq:6p}
 \dot \sigma/k =  \, \frac{\mathsf{\Lambda}_\rho  \ell_R \, \dot \rho +\mathsf{\Lambda} \dot \ell_R}{1 - k \mathsf{\Lambda}_\sigma  \ell_R },
  \qquad 
 \ee
where we have set  $\mathsf{\Lambda}_\sigma = \partial \mathsf{\Lambda}/ \partial \sigma$ and $\mathsf{\Lambda}_\rho = \partial \mathsf{\Lambda}/ \partial \rho$. 
  However from \eqref{eq:1p}$_1$ and \eqref{eq:1p}$_4$
$$
\ell_R = \frac{\ell}{\mathsf{\Lambda}} {=} \frac{k\ell_0 + \sigma}{k\mathsf{\Lambda}} {=} \frac{\sigma_{0} + \sigma}{k\mathsf{\Lambda}}.
$$
Using this to eliminate $\ell_R$ from  \eqref{eq:6p} yields
 \be
 \label{eq:8p}
\frac{\dot \sigma}{\sigma_0} =  
\, \frac{ \mathsf{\Lambda}  \dot\ell_R/\ell_0 +  (1 +  \sigma /\sigma_{0} )\dot \rho  \mathsf{\Lambda}_\rho/\mathsf{\Lambda} 
  }
{1 -   ( \sigma + \sigma_{0} )  \mathsf{\Lambda}_\sigma/\mathsf{\Lambda}}. \qquad
 \ee

Equation \eqref{eq:1p}$_3$ can be solved for $\rho(t)$.  When this expression for $\rho(t)$, together with $\mathsf{\Lambda} = \mathsf{\Lambda}(\sigma, \rho)$ and $\dot \ell_R = \mathsf{V}(\sigma, \rho)$, are substituted into \eqref{eq:8p}, the resulting equation has the form $\dot \sigma = F(\sigma, t)$.  This can be solved for $\sigma(t)$. Thereafter one can calculate the associated value of $\dot\ell(t)$ from 
\be
\label{eq:20200108-3}
{\dot\ell}/{\ell_o} \stackrel{\eqref{eq:4p}}{=} {\dot \sigma}/{(k\ell_0)} \stackrel{\eqref{eq-0501-sig0}}{=}  {\dot \sigma}/{\sigma_{0}} ,\qquad 
\ee
and thus one can construct a parametric plot of $(\sigma(t),\dot\ell(t))$ on the $\sigma,\dot\ell$-plane with time $t$ as the parameter. The resulting figure can be compared with Figures 2B and 2C of Parekh {\it et al.}   \cite{ParekhChaudhuri2005}.

\begin{figure}[ht!]
\centering
\includegraphics[scale=0.55]{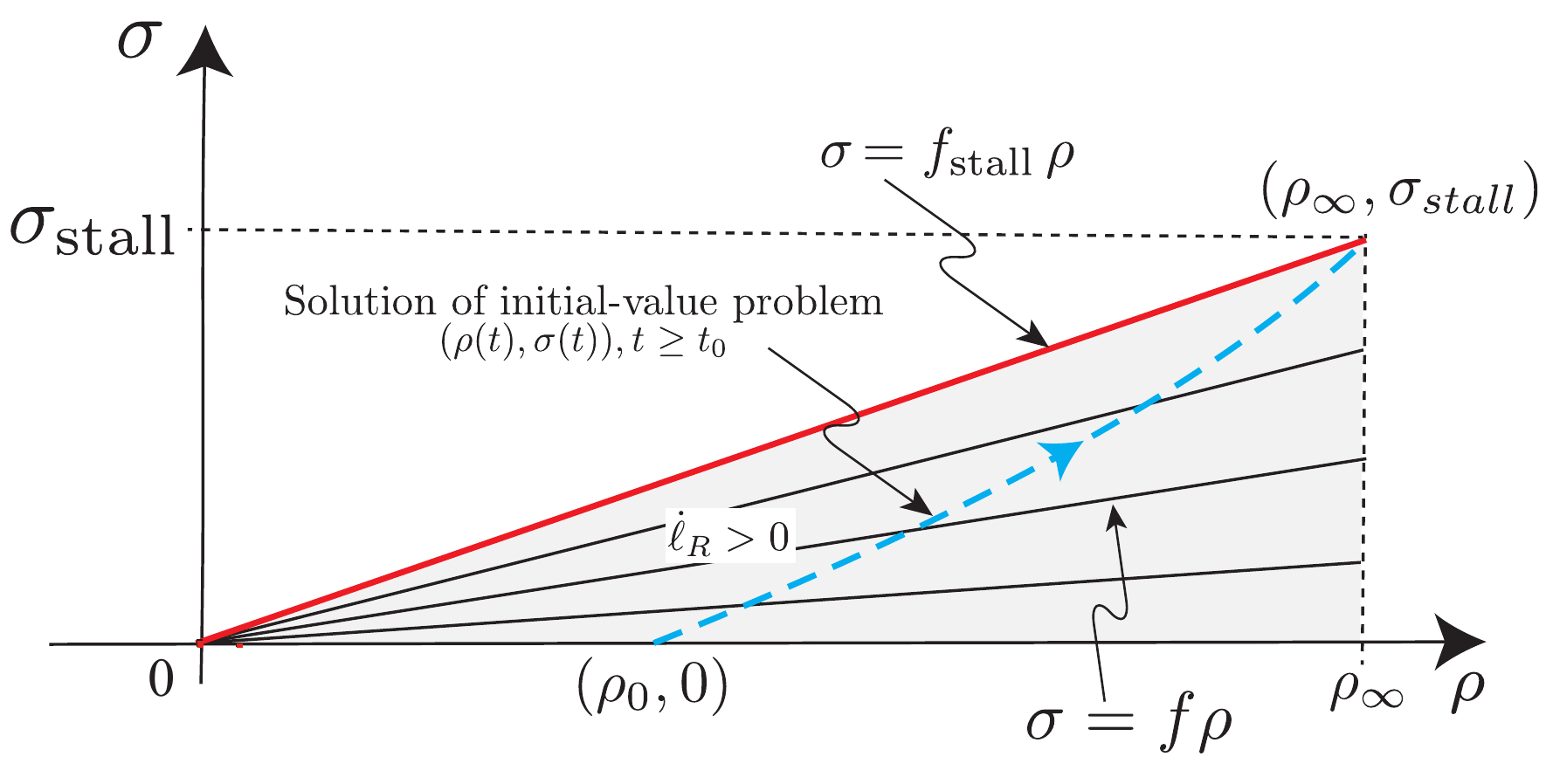}
\caption{The radial straight lines $\sigma = f \rho$ are lines of constant filament force. The kinetic relation $V = \mathsf{V}(f)$ yields $V = 0$ on the bold red line $\sigma = f_{stall}\, \rho$, and  $V >0$ on the shaded region below it. The trajectory defined by a solution $(\rho(t), \sigma(t)), t \geq t_0,$ of a generic initial-value problem is depicted schematically by the blue dashed curve. It starts at the initial point $(\rho_0, 0)$ and terminates at stall corresponding to $(\rho_\infty, \sigma_{stall})$.}
\label{Fig-rho-sig-plane0}
\end{figure}

Before proceeding to do this, since it will be useful when we discuss the dissipation inequality later on, we now turn briefly to  the $\rho, \sigma$-plane.  Each radial line $\sigma = f \rho$ on this plane corresponds to a constant filament force $f$; see Figure \ref{Fig-rho-sig-plane0}. We are interested in the range $0 \leq f \leq f_{stall}, 0 \leq \rho \leq \rho_\infty,$ corresponding to the 
wedge-shaped shaded region in the figure.  The kinetic relation for surface growth has the form $V = \mathsf{V}(f)$ and so each radial line also corresponds to a constant growth speed $V$.  Since $\mathsf{V}(f_{stall})=0$ and $\mathsf{V}(f_{stall}) > 0$ for $0\leq f < f_{stall}$, the growth speed  vanishes on the bold red line $\sigma = f_{stall}\, \rho$ and is positive below it.  Now consider a generic initial-value problem for a spring-loaded specimen. In this case one solves the differential equations $\dot \sigma = \mathsf{S}(\sigma, \rho), \dot \rho = \mathsf{R}(\rho)$ subject to initial conditions, say $\sigma(t_0) = 0, \rho(t_0) = \rho_0$, where $\mathsf{S}$ is given by  \eqref{eq:8p},  \eqref{eq:1p}$_{2,3}$ and $\mathsf{R}$ by \eqref{eq:20191219-1}. The solution $\sigma = \sigma(t), \rho = \rho(t), t \geq t_0,$ of this problem describes a trajectory in the $\rho,\sigma$-plane shown schematically by the dashed blue curve in Figure  \ref{Fig-rho-sig-plane0}. It starts at $(\rho, \sigma) = (\rho_0, 0)$ and terminates at $(\rho, \sigma) = (\rho_\infty, \sigma_{stall})$ corresponding to stall.

%


\subsubsection{Simple $\rho$-independent model.}

Before presenting the results of the preceding analysis, it is illuminating to first consider a simpler model in which  the number of filaments is fixed and does not evolve. Then we drop $\rho$ from the model\footnote{It should be noted that this is not the same model we used in Section \ref{sec-3} since here we are using the regularized maximum dissipation kinetic relation for surface growth. This is needed in order to approach stall gradually.}  so that equations \eqref{eq:8p} and \eqref{eq:20200108-3} governing $\ell(t), \sigma(t)$ specialize to
\be
\label{eq-20200131-2}
\dot\ell = \frac{\mathsf{\Lambda}(\sigma) \mathsf{V}(\sigma) }{1 - (\sigma + \sigma_{0})\mathsf{\Lambda}'(\sigma)/\mathsf{\Lambda}(\sigma)}, \qquad \frac{\dot\sigma}{\sigma_{0}} = \frac{\dot \ell}{\ell_0} , 
\ee
where we have set $V = V(\sigma)$. 
Equation \eqref{eq-20200131-2}$_1$ gives $\dot\ell$ as a function of stress $\sigma$, which can be plotted on the $\sigma, \dot\ell$-plane.  Moreover, integrating \eqref{eq-20200131-2} from $\sigma=0$ at $t=t_0$ to $\sigma = \sigma_{stall}$ at $t=t_{stall}$ gives the time $t_{stall}$ at which growth stalls.

\begin{figure}[h]
\centerline{\includegraphics[scale=0.25]{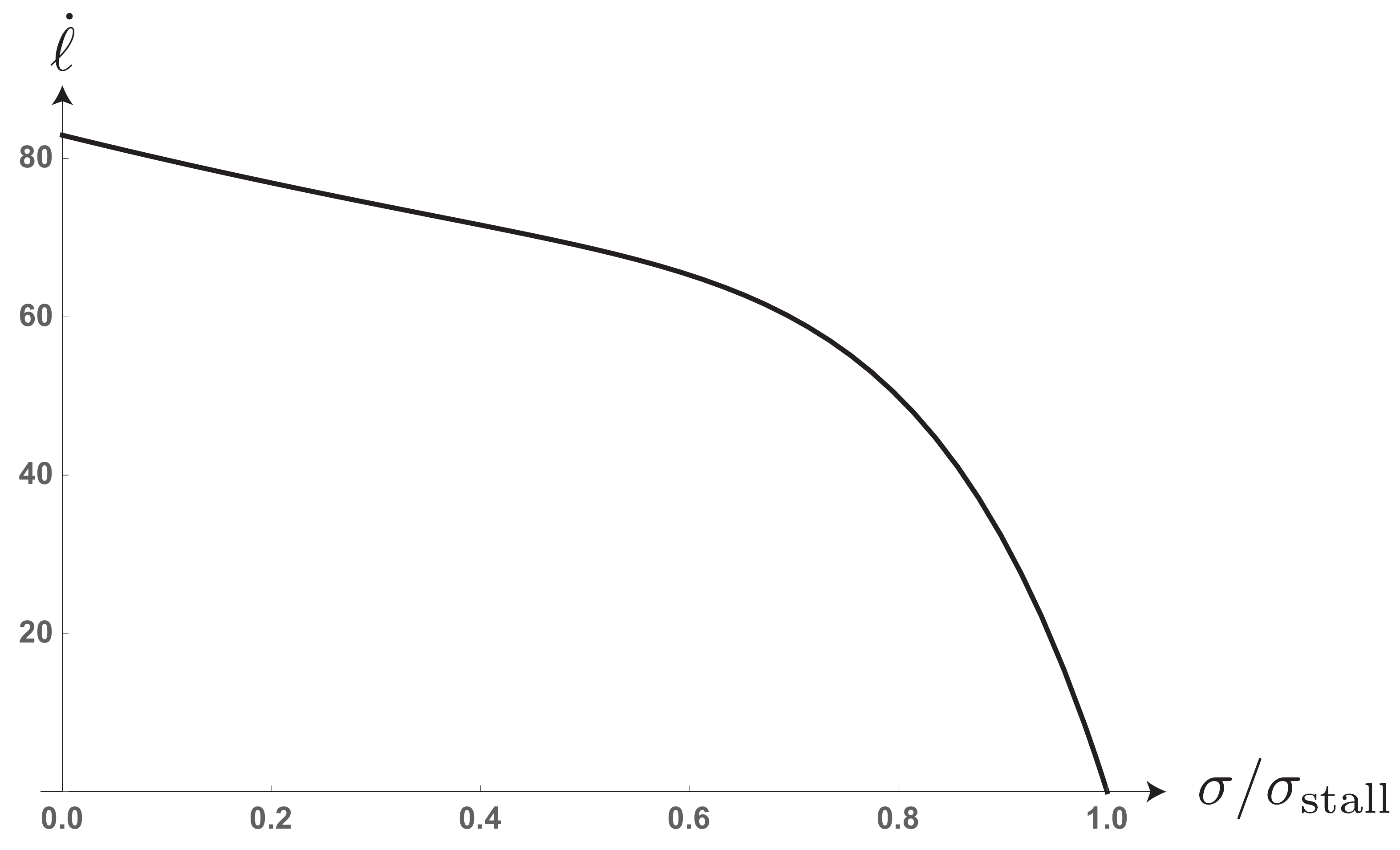}}

\caption{{Elongation-rate $\dot\ell$ versus stress $\sigma$ according to the simple model \eqref{eq-20200131-2}, \eqref{eq-20200131-4}.}}

\label{Fig-Parekh-Simple}
\end{figure}

We now specialize \eqref{eq-20200131-2} to the particular choices  \eqref{eq:20200125-1}, \eqref{eq:20191206-3},
\be
\label{eq-20200131-4}
\mathsf{\Lambda}(\sigma) = 1/(1 +\sigma/E), \qquad \mathsf{V}(\sigma) = V_0\left[ 1 - \left(\frac{\sigma}{\sigma_{stall}}\right)^m \right],
\ee
with the various parameters having the values given in \eqref{eq:20200208-45} and  \eqref{eq:20200208-46}.
Figure \ref{Fig-Parekh-Simple} shows the plot of $\dot\ell$ versus $\sigma$ according to this model.  Observe that the elongation-rate starts at $\dot \ell = 83 \, \si{nm/min}$ and eventually stalls when the force reaches the value $\sigma_{stall} A = 293\, \si{nN}$.  The time taken for the stress to reach $99\%$ of $\sigma_{stall}$ is $235\, \si{min}$.  This figure may  be compared with Figures 2B and 2C of Parekh {\it et al.}  \cite{ParekhChaudhuri2005} where the elongation-rate starts at about $72 \, \si{nm/min}$ and growth stalls at a force of about $300\, \si{nN}$ in a little over $200\, \si{min}$.

\subsubsection{General $\rho$-dependent model.}

\begin{figure}[h]
\centerline{\includegraphics[scale=0.30]{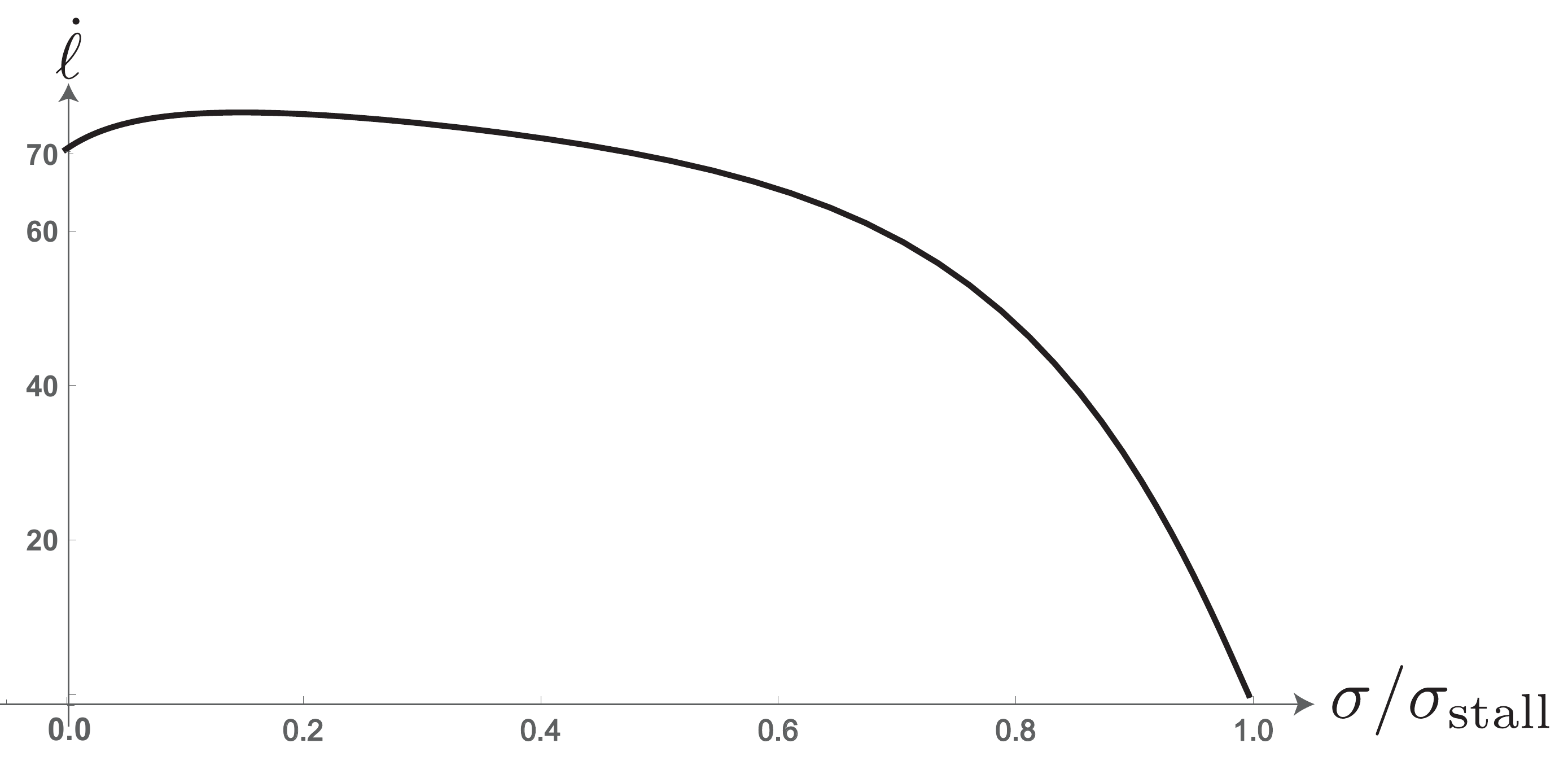}}

\caption{{Elongation-rate $\dot\ell$ versus stress $\sigma$ according to the general model \eqref{eq:8p}, \eqref{eq:20200108-3}.}}

\label{Fig-20200203-SpringComplex}
\end{figure}

Now consider the more detailed model  \eqref{eq:8p}, \eqref{eq:20200108-3}, specialized to the constitutive descriptions \eqref{eq:20200125-1}, \eqref{eq:20191219-1}, \eqref{eq:20191206-3}.  The parameters have the same values used in the preceding subsection together with $\tau_\rho$ taking the value given in \eqref{eq:20200208-45}$_3$. The initial condition for $\rho$ is chosen (arbitrarily) to be $\rho(t_0)=0.5\rho_\infty$.
Figure \ref{Fig-20200203-SpringComplex} shows a parametric plot of $(\sigma(t),\dot\ell(t))$ on the stress-elongation-rate plane with time being the parameter.  The elongation-rate starts at the value $70\, \si{nm/min}$ and rises to a maximum value of $75\, \si{nm/min}$.  The elongation-rate remains at about $70\, \si{nm/min}$ during a more-or-less load-independent intermediate stage after which $\dot\ell$ begins to decrease more rapidly as the system approaches stall.  The stress increases monotonically throughout this calculation until it  reaches\footnote{According to the kinetic law \eqref{eq:20191219-1} it takes infinite time for $\rho(t)$ to reach the value $\rho_\infty$ which is why we calculate the time to reach $99\%$ of stall.}  $99\%$ of the stall force in $200\, \si{min}$, the stall force being $293\, \si{nN}$.
As noted above, the corresponding experimental values from \cite{ParekhChaudhuri2005} are an initial elongation-rate of about $72 \, \si{nm/min}$, growth stalling at a force of about $300\, \si{nN}$ in a little over $200\, \si{min}$.

%


 \subsection{Stress jumps.} \label{sec-0505-13}
 
We now turn to two calculations motivated by Figures 3a and 3b of Parekh {\it et al.} \cite{ParekhChaudhuri2005} where the force on the specimen is {\it suddenly} decreased while it is growing.  Our primary interest is in calculating the resulting jump in the elongation-rate $\dot\ell$ which, the experiments indicate, undergoes a striking sudden increase.

It is worth noting at the outset that the kinetic law $\dot \rho = \mathsf{R}(\rho)$ is independent of stress and so is unaffected by the details of how  the stress varies.  Thus we take $\rho(t)$ and $\dot \rho(t)$ to vary continuously throughout. Consequently  the effective Young's modulus, $E = \rho^2 E_\infty/\rho_\infty^2$, is also a continuous function of time.  Moreover, keeping in mind that the referential length of the specimen $\ell_R(t)$ changes {\it only} due to growth, and assuming that a finite segment of new material cannot appear in an infinitesimal instant of time, we  require $\ell_R(t)$ to be a continuous function of time.  On the other hand the specimen length $\ell(t)$ and elongation-rate $\dot\ell(t)$ will be discontinuous when the stress is discontinuous.

In the {\it first} calculation the specimen grows under spring loading conditions during an initial period $t_0 < t < t_2$; see left-hand figure in Figure \ref{Fig-3a.pdf}.  At time $t_2$, the force is suddenly decreased to a smaller value and clamped at that value from then on. (The specimen is {\it not} under spring-loading conditions for $t > t_2$.)

In the {\it second} calculation  (see left-hand figure in Figure \ref{Fig-3b.pdf}) the specimen grows with the force clamped at a fixed value during an initial stage $t_0 < t < t_1$.  At the instant $t=t_1$ the force clamp is released and the specimen is allowed to grow under spring loading conditions for a period $t_1 < t < t_2$.  At the instant $t=t_2$ the force is suddenly decreased back to the value it had during the original constant force stage and clamped at that value for $t > t_2$.

In both of these calculations the load levels are always in the ``load-independent'' range of Figure \ref{Fig-20200203-SpringComplex}.

In all processes, whether the stress is held constant or the specimen is spring-loaded, we have
\be
\label{eq:20200129-1}
\ell = \mathsf{\Lambda} \ell_R, \qquad \dot \ell_R = \mathsf{V}(\sigma,\rho), \qquad \dot\rho = \mathsf{R}(\rho). 
\ee
Differentiating \eqref{eq:20200129-1}$_1$ with respect to time, and then using \eqref{eq:20200129-1}$_1$ and \eqref{eq:20200129-1}$_2$ to eliminate $\ell_R$ and $\dot\ell_R$ from the result, leads to
$$
\dot\ell =  \mathsf{\Lambda} \mathsf{V} + \ell \, \frac{\mathsf{\Lambda}_\sigma}{\mathsf{\Lambda}}\, \dot \sigma  + \ell \, \frac{\mathsf{\Lambda}_\rho}{\mathsf{\Lambda}} \dot \rho . 
$$
In processes where $\sigma(t) = \rm constant$, this reduces to
\be
\label{eq:20200129-3}
\mbox{Constant stress:} \qquad \dot\ell =  \mathsf{\Lambda} \mathsf{V}  + \ell \frac{\mathsf{\Lambda}_\rho}{\mathsf{\Lambda}} \dot \rho , 
\ee
whereas when the specimen is spring-loaded, so that $\sigma = k(\ell - \ell_0)$, it yields
\be
\label{eq:20200129-4}
\mbox{Spring loading:} \qquad\dot\ell = \frac{\mathsf{\Lambda} \mathsf{V} +   \dot\rho \, \ell_0(1 + \sigma/\sigma_0) \,  \mathsf{\Lambda}_\rho /\mathsf{\Lambda} }{1 - (\sigma + \sigma_{0}) \mathsf{\Lambda}_\sigma/\mathsf{\Lambda}},  
\ee
where $\sigma_{0} = k \ell_0$ as before.

Consider the instant $t_2$ at which the stress changes discontinuously. For any time-dependent function $g(t)$ that suffers a finite jump discontinuity at time $t_2$ we write
$$
g^+ = g(t_2^+), \qquad g^- = g(t_2^-),  
$$ 
and if $g(t)$ is continuous at $t_2$ we simply write $g = g(t_2)$.  In both of Parekh {\it et al.}'s experiments, the specimen is spring-loaded just before the stress jump and has the force clamped just after. Thus \eqref{eq:20200129-4} holds at $t_2^-$, while \eqref{eq:20200129-3} holds at $t_2^+$.  The elongation-rates just before and just after the stress jump are therefore
\be
\label{eq:0125-13x}
\dot\ell^- = \frac{\mathsf{\Lambda}^- \mathsf{V}^- +   \dot\rho \, \ell_0(1 + \sigma^-/\sigma_0) \mathsf{\Lambda}^-_\rho /\mathsf{\Lambda}^- }{1 - (\sigma^- + \sigma_{0}) \mathsf{\Lambda}^-_\sigma/\mathsf{\Lambda}^-}, \qquad
\dot\ell^+ =  \mathsf{\Lambda}^+ \mathsf{V}^+  +
 \dot \rho \ell^+  {\mathsf{\Lambda}^+_\rho}/{\mathsf{\Lambda}^+} , 
\ee
where we have written $\mathsf{\Lambda}^\pm = \mathsf{\Lambda}(\sigma^\pm, \rho)$ and $\mathsf{V}^\pm =\mathsf{V}(\sigma^\pm,\rho)$ having used the fact that
 $\rho(t)$ varies continuously.

Consider an instant at which the stress has the values $\sigma^\pm$.  In order to calculate $\dot\ell^+$ using \eqref{eq:0125-13x}$_2$ we need the value of $\ell^+$.   While we can use $\sigma^- = k (\ell^- -\ell_0)$ to calculate $\ell^-$, we cannot use  $\sigma^+ = k (\ell^+ -\ell_0)$ to calculate $\ell^+$ since the specimen is not spring-loaded at time $t^+$.  Instead, we calculate $\ell^+$ using  $\ell/\ell_R = \mathsf{\Lambda}$, i.e. 
\be
\label{eq:0122-26x}
\ell^+ = \frac{\mathsf{\Lambda}(\sigma^+,\rho)}{\mathsf{\Lambda}(\sigma^-,\rho)} \, \ell^-. 
\ee
In writing \eqref{eq:0122-26x} we have used the fact noted at the beginning of this section that  the filament density $\rho(t)$ and referential length $\ell_R(t)$ vary continuously.


\subsubsection{Simple $\rho$-independent model.} \label{subsec-0505-131}

In order to get a sense for how the elongation-rate suffers a sudden increase when the stress is suddenly decreased, consider again the special case where the constitutive functions $\mathsf{\Lambda}$ and $\mathsf{V}$ are both independent of the filament density $\rho$. 
Then equations \eqref{eq:20200129-3} and \eqref{eq:20200129-4} take the forms
\be
\label{eq:0120-1}
\mbox{Constant stress:} \qquad  \dot \ell = \mathsf{\Lambda}(\sigma) \mathsf{V}(\sigma), 
\ee
\be
\label{eq:0120-2}
\mbox{Spring loading:} \qquad  \dot\ell = \frac{ \mathsf{\Lambda}(\sigma) \mathsf{V}(\sigma)}{1 - (\sigma + \sigma_{0})\mathsf{\Lambda}'(\sigma)/\mathsf{\Lambda}(\sigma)}. 
\ee
These equations tell us how the elongation-rate $\dot\ell$ varies as a function of stress $\sigma$ for the two types of loading.

\begin{figure}[h]
\centerline{\includegraphics[scale=0.30]{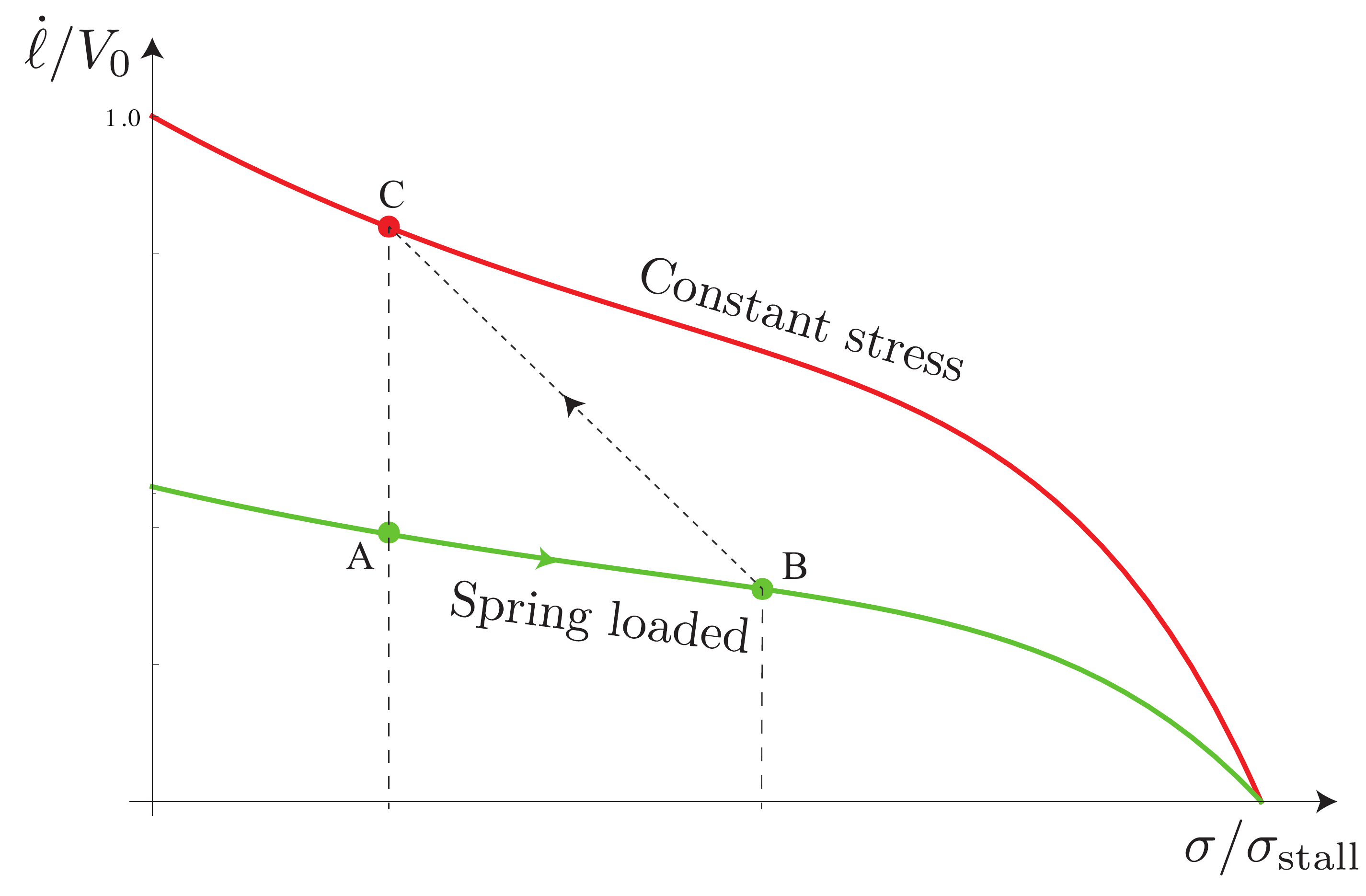}}
\caption{{{Elongation-rate $\dot\ell$ versus stress $\sigma$ for (a) loading at constant stress (equation \eqref{eq:0120-1}, red curve) and (b) spring-loading (equation \eqref{eq:0120-2}, green curve). Suppose  that the spring-loaded specimen evolves from A to B, at which point the stress is suddenly decreased back to the value $\sigma(A)$ and clamped at that value. Accordingly the system must jump from B to C and remain there from then on. Thus during the jump, the stress decreases discontinuously from $\sigma(B)$ to $\sigma(C)\, (= \sigma(A))$ while the elongation-rate increases discontinuously from $\dot\ell(B)$ to $\dot\ell(C)$.}}}
\label{fig-Responses.pdf}
\end{figure}

Subject to mild assumptions on $\mathsf{\Lambda}$ and $\mathsf{V}$,  the curve on the $\sigma, \dot\ell$-plane defined by \eqref{eq:0120-2} lies below the one described by \eqref{eq:0120-1}. 
For example, Figure \ref{fig-Responses.pdf} shows these curves for the particular choice  \eqref{eq-20200131-4}
with the parameters having the values given in \eqref{eq:20200208-45} and \eqref{eq:20200208-47}.

When $\sigma=0$ we have $\mathsf{\Lambda}(0)=1, \mathsf{\Lambda}'(0)=-1/E$ and $\mathsf{V}(0)=V_0$, and therefore from \eqref{eq:0120-1} and \eqref{eq:0120-2} 
$$
\dot \ell = \left\{
\ba{cll}
V_0 \qquad &\mbox{(constant force),}\\[2ex]
 \displaystyle \frac{V_0}{1 + \sigma_{0}/E} \qquad &\mbox{(spring loading),} \\
\ea \qquad \qquad {\rm at} \ \sigma = 0. 
\right.
$$
These are the values of $\dot\ell$ at which the curves in Figure \ref{fig-Responses.pdf} intersect the vertical axis.  Therefore the separation between the two curves (at least at $\sigma=0$) increases as $\sigma_0/E$ increases.
Therefore in order to increase the separation at $\sigma =0$ we should decrease $E$. In our quantitative calculations we have therefore taken the smallest value of $E$ from the range of possible values determined experimentally. The value of $\sigma_0$ is determined by the stiffness of the AFM spring.

Now consider the response of a spring-loaded specimen starting from point $A$  in 
Figure \ref{fig-Responses.pdf} where the stress is $\sigma(A)$. The point $(\sigma(t), \dot\ell(t))$ evolves along the green curve starting from $A$ and moving to the right (towards stall).  Suppose that when it reaches point $B$ the stress is suddenly decreased back to the value $\sigma(A)$ {\it and clamped} at that value. Then  $(\sigma(t), \dot\ell(t))$ jumps from point $B$ to point $C$ (and remains there for subsequent time).  Therefore, as the stress decreases suddenly from $\sigma(B)$ to $\sigma(C) (= \sigma(A))$ the elongation-rate increases discontinuously from the value $\dot \ell(B)$ to $\dot \ell(C)$.

In our {\it first}\footnote{First and second refer to the two processes described at the very beginning of Section \ref{sec-4}\ref{sec-0505-13}.}  quantitative calculation we took the values of the force just before and after the jump from Figure 3a of \cite{ParekhChaudhuri2005} and determined the elongation-rates from
\be
\label{eq:0505-1}
\dot\ell^- = \frac{ \mathsf{\Lambda}(\sigma^-) \mathsf{V}(\sigma^-)}{1 - (\sigma^- + \sigma_{0})\mathsf{\Lambda}'(\sigma^-)/\mathsf{\Lambda}(\sigma^-)}, \qquad  \dot \ell^+ = \mathsf{\Lambda}(\sigma^+) \mathsf{V}(\sigma^+),  
\ee
which follow from  \eqref{eq:0120-1} and \eqref{eq:0120-2}.  The functions $\mathsf{\Lambda}$ and $\mathsf{V}$ are given by \eqref{eq-20200131-4}  with the parameters having the values in  \eqref{eq:20200208-45} and \eqref{eq:20200208-47}.  Equation \eqref{eq:0505-1} then led to 
$$
\dot \ell^- = 120 \, \si{nm/min}, \qquad  \dot \ell^+ = 204\, \si{nm/min}.
$$
The corresponding experimentally measured elongation-rates were $\dot\ell^-=129 \, \si{nm/min}$ and $\dot\ell^+=275 \, \si{nm/min}$.

In our {\it second} numerical calculation the values of the force just before and after the jump were taken from Figure 3b of  \cite{ParekhChaudhuri2005}. The  elongation-rates were calculated as above and this led to
$$
\dot \ell^- = 162 \, \si{nm/min}, \qquad  \dot \ell^+ = 239\, \si{nm/min},
$$
whereas the experimentally measured values were $\dot\ell^-=129 \, \si{nm/min}$ and $\dot\ell^+=270 \, \si{nm/min}$.


\subsubsection{General $\rho$-dependent model.} \label{subsec-0505-132}

When the filament density $\rho$ is taken into account, a simple graphical description based on \eqref{eq:20200129-3} and \eqref{eq:20200129-4} is no longer possible since the right-hand sides of those equations now depend on both $\sigma$ and $t$ (through $\rho(t)$).  Instead we solve the relevant initial-value problem based on \eqref{eq:20200129-3} and \eqref{eq:20200129-4} to calculate the response of the specimen and in particular to determine the elongation-rates just before and after the stress jump.

 \begin{figure}[h]
\begin{minipage}{0.45\linewidth}
\centerline{\includegraphics[scale=0.25]{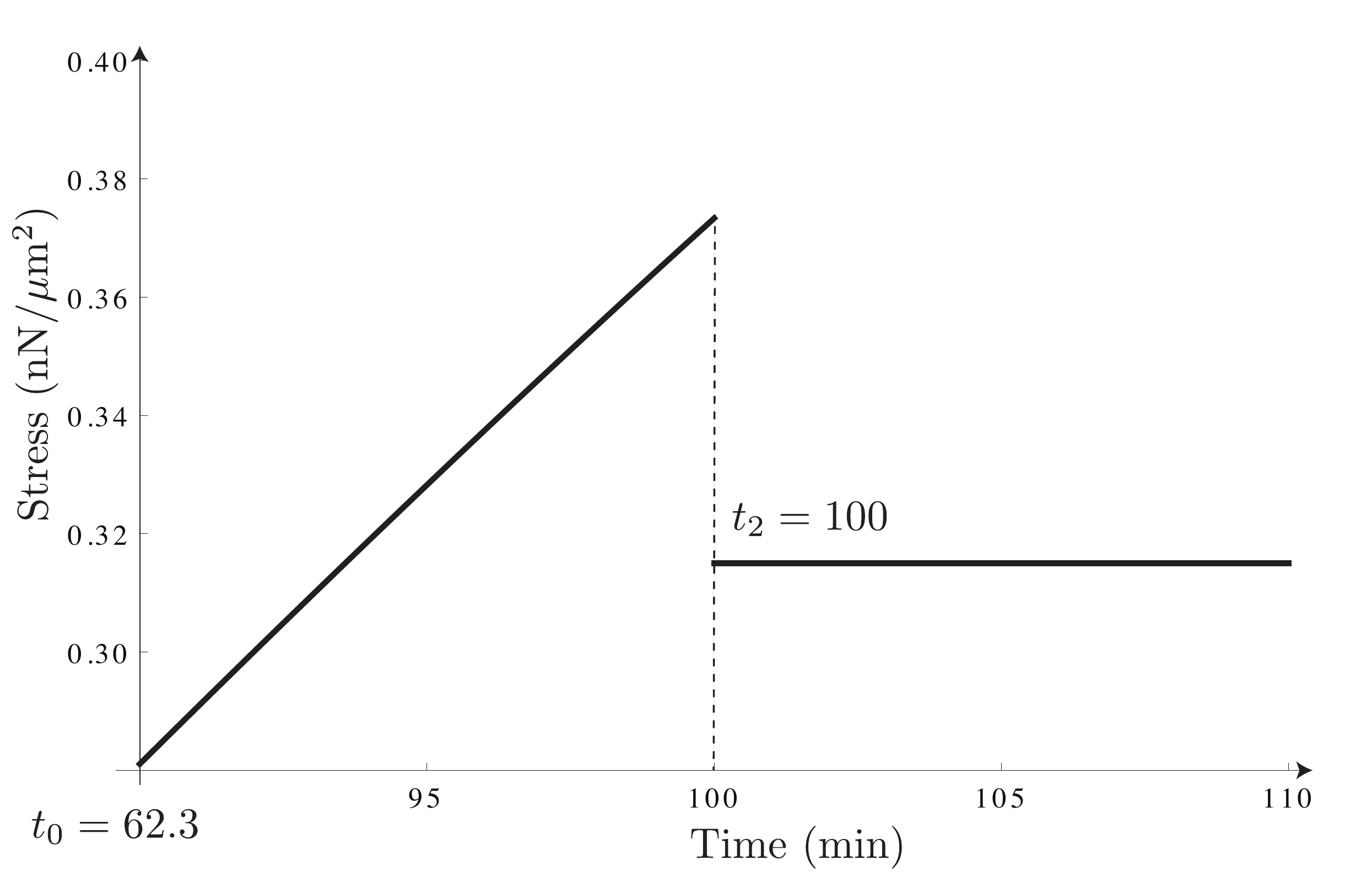}}
\end{minipage}
\hspace{0.5cm}
\begin{minipage}{0.45\linewidth}
\centerline{\includegraphics[scale=0.25]{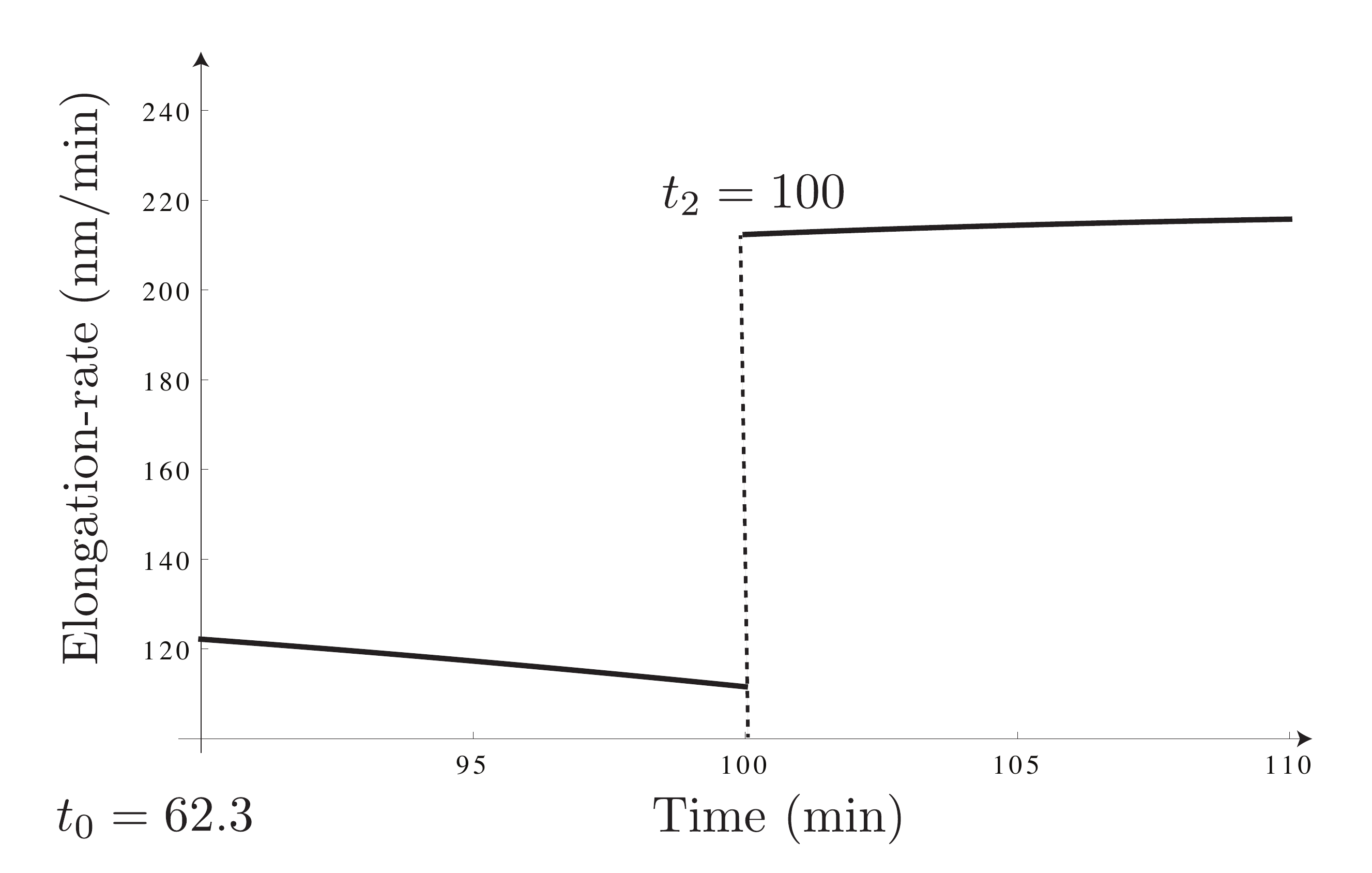}}
\end{minipage}
\caption{{\footnotesize Stress and elongation-rate versus time; compare with Figure 3a of \cite{ParekhChaudhuri2005}. The specimen grows under spring loading for $t_0 < t < t_2$ and with the stress fixed for $t>t_2$.}}
\label{Fig-3a.pdf}
\end{figure}

In the {\it first} calculation we solve the differential equation 
\eqref{eq:20200129-4} for $t_0 < t < t_2$ using suitable initial conditions at $t_0$.  The conditions at time $t_2^+$ are then determined using \eqref{eq:0122-26x} and this information is used as initial conditions to solve \eqref{eq:20200129-3} for $t > t_2$. The details of this calculation can be found in Section S3 of the  Supplementary Material. Figure \ref{Fig-3a.pdf}  shows plots of $\sigma(t)$ and $\dot\ell(t)$ versus time as predicted by our model which may be compared with Figure 3a of \cite{ParekhChaudhuri2005}. In particular we find $\sigma(t_2^-) = 0.373\, \si{nN/\mu m^2}, \dot\ell(t_2^-) = 112\, \si{nm/min}$ and $\dot\ell(t_2^+) = 212\, \si{nm/min}$, the corresponding experimentally determined values  being 
$\sigma(t_2^-) = 0.381\, \si{nN/\mu m^2}$, $\dot\ell(t_2^-) =129 \, \si{nm/min}$ and $\dot\ell(t_2^+)=275 \, \si{nm/min}$.

 \begin{figure}[h]
\begin{minipage}{0.45\linewidth}
\centerline{\includegraphics[scale=0.32]{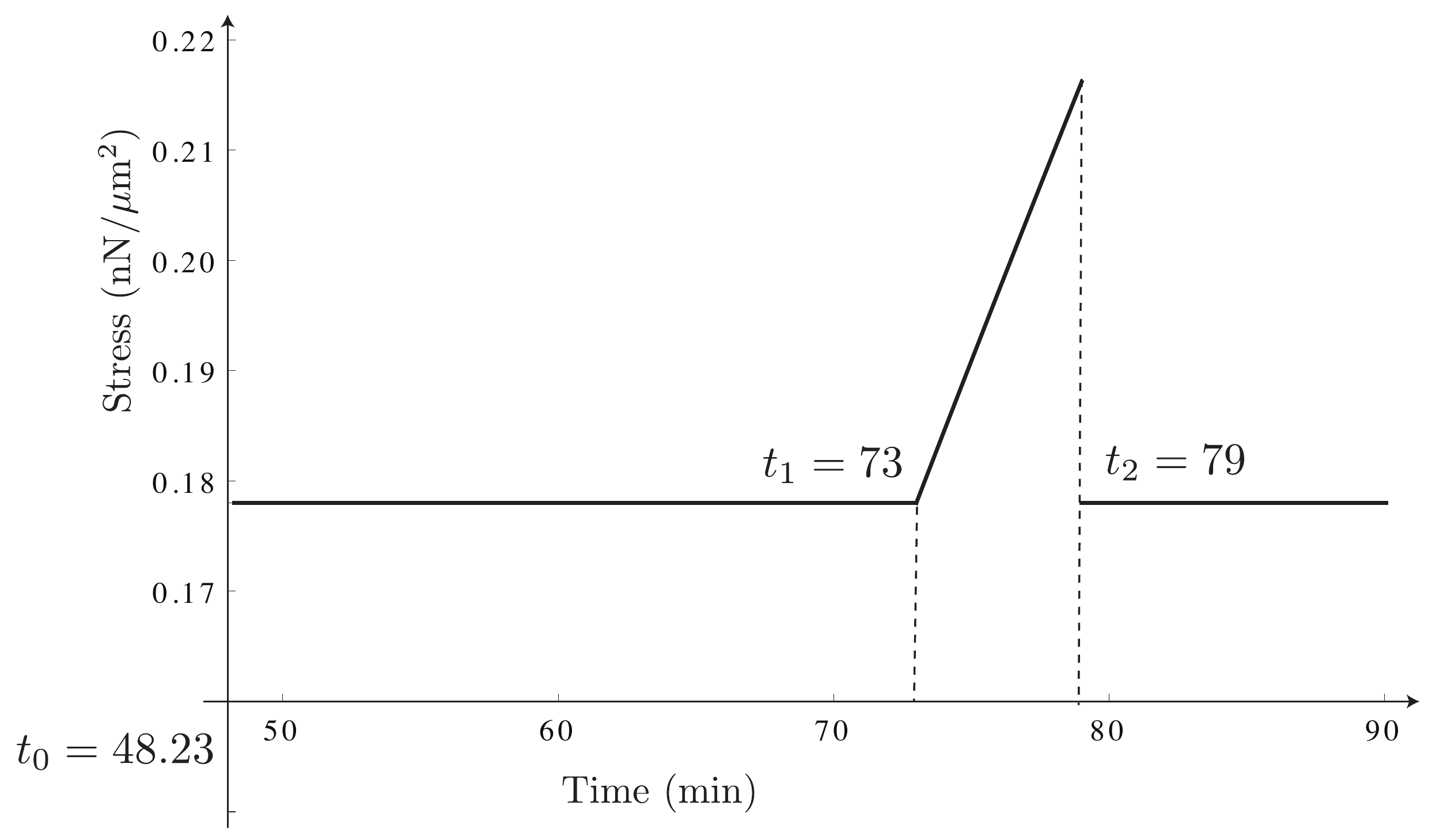}}
\end{minipage}
\hspace{0.5cm}
\begin{minipage}{0.45\linewidth}
\centerline{\includegraphics[scale=0.32]{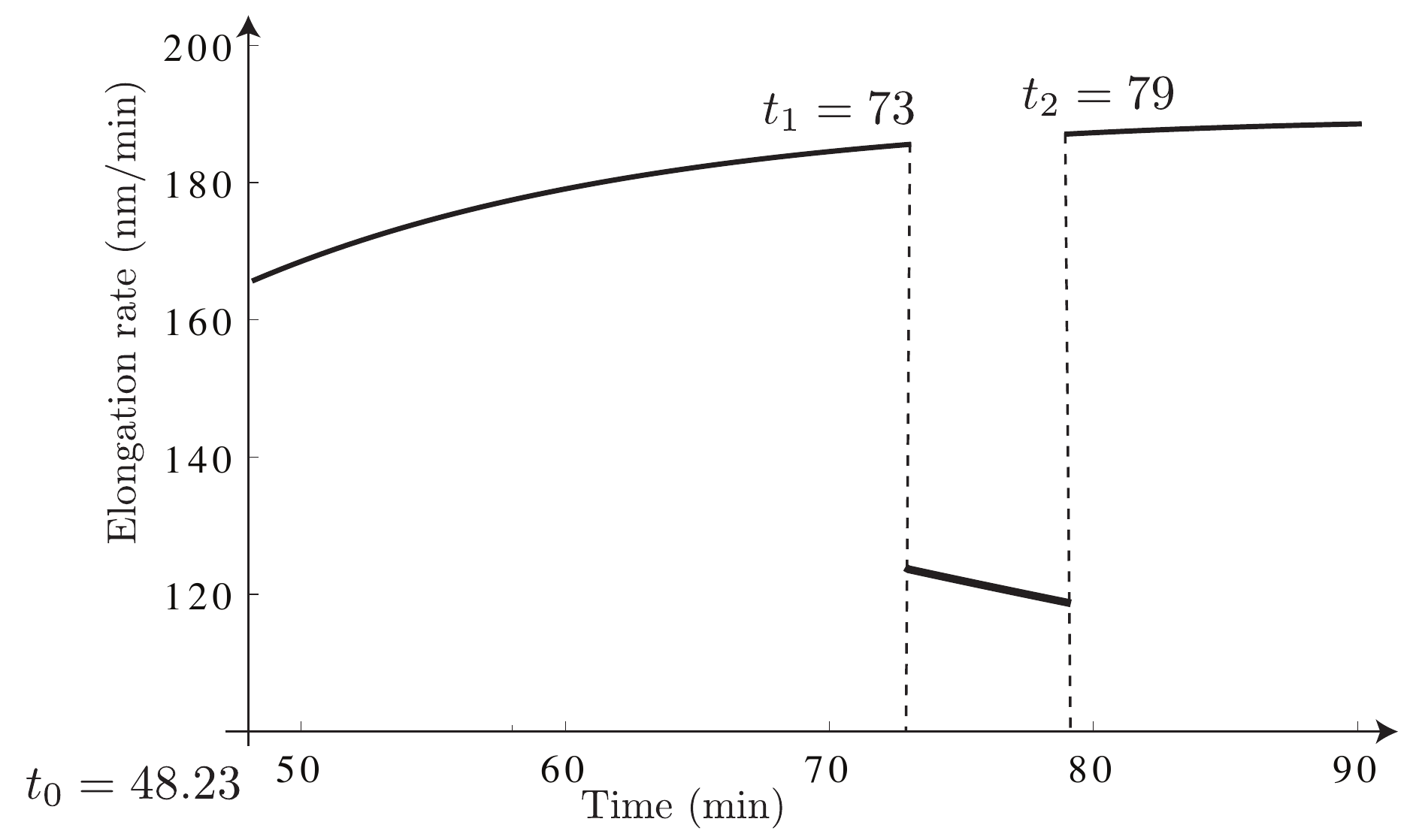}}
\end{minipage}
\caption{{\footnotesize Stress and elongation-rate versus time; compare with Figure 3b of \cite{ParekhChaudhuri2005}. The specimen grows with the stress fixed at the same value for both $t_0 < t < t_1$ and $t > t_2$, and under spring loading for intermediate times $t_1 < t < t_2$.}}
\label{Fig-3b.pdf}
\end{figure}

In the {\it second} calculation  we start by solving \eqref{eq:20200129-3} for $t_0 < t < t_1$ using suitable initial conditions at $t_0$. The conditions at time $t_1^+$ are then deduced from continuity and the results are used as initial conditions to solve \eqref{eq:20200129-4} for $t_1 < t < t_2$. The conditions at time $t_2^+$ are then calculated using \eqref{eq:0122-26x}. Finally we solve \eqref{eq:20200129-3} for $t > t_2$ using the information from $t_2^+$ as initial conditions. The details of these calculations can be found in Section S3 of the  Supplementary Material.  Figure \ref{Fig-3b.pdf}  shows plots of $\sigma(t)$ and $\dot\ell(t)$ versus time as predicted by our model which may be compared with Figure 3b of \cite{ParekhChaudhuri2005}. In particular we find $\sigma(t_2^-) = 0.216\, \si{nN/\mu m^2}, \dot\ell(t_2^-) = 119\, \si{nm/min}$ and $\dot\ell(t_2^+) = 187\, \si{nm/min}$, the corresponding experimentally determined values  being 
$\sigma(t_2^-) = 0.218\, \si{nN/\mu m^2}$, $\dot\ell(t_2^-) =129 \, \si{nm/min}$ and $\dot\ell(t_2^+)=270 \, \si{nm/min}$.

In this second calculation the stress is held fixed at the same value $0.218\, \si{nN/\mu m^2}$ for both $t_0 < t < t_1$ and $t >t_2$.   Therefore on both these time intervals the corresponding elongation-rate is given by \eqref{eq:20200129-3} with $\sigma = 0.218\, \si{nN/\mu m^2}$.  However  the right-hand side of \eqref{eq:20200129-3} also involves the filament density $\rho(t)$ and therefore though $\sigma$ has the same constant value, the elongation-rate evolves as a function of time due to the evolution of $\rho(t)$. In the specific calculation above, we find (in particular) that $\rho(t_1) = 0.838\rho_\infty$ and $\rho(t_2)= 0.861\rho_\infty$ at $t_1 = 73\, \si{min}$ and $t_2 = 79\, \si{min}$. This small difference in the filament densities leads to a small difference in the  corresponding elongation-rates, viz. $\dot\ell(t_1^-) =186 \, \si{nm/min}$ and $\dot\ell(t_2^+) = 187\, \si{nm/min}$. This is in contrast to the large difference, $\dot\ell(t_1^-) =170 \, \si{nm/min}$ and $\dot\ell(t_2^+)=270 \, \si{nm/min}$, observed in the experiments.
In order to capture this we will need to modify the kinetic relation $\dot\rho = \mathsf{R}(\rho)$ and possibly $\dot\ell_R = \mathsf{V}(\sigma, \rho)$ and $\lambda = \mathsf{\Lambda}(\sigma, \rho)$ as well.

Finally it is worth noting the qualitative similarity between the right-hand figure in Figure \ref{Fig-3b.pdf} and  Figure \ref{Fig-ell-t-jumps.pdf} from our earlier calculation related to the experiments of Brangbour {\it et al.} \cite{brangbour}. In both cases the force on the specimen is held constant at the same value for $t_0 < t< t_1$ and $t>t_2$. In the case of Figure \ref{Fig-ell-t-jumps.pdf} the force was fixed at a smaller value during the intermediate interval $t_1 < t < t_2$ whereas the specimen grew under spring loading in Figure \ref{Fig-3b.pdf}.  In Figure \ref{Fig-ell-t-jumps.pdf}, the curve (straight-line) pertaining to $t>t_2$ is the continuation of the curve pertaining to $t_0 < t < t_1$. Because of the dependency on the filament density $\rho(t)$, this is not true of the corresponding curves in Figure \ref{Fig-3b.pdf}.



\section{Conclusion}
In this paper we have shown that a continuum model of surface growth and filament nucleation can quantitatively capture the evolution of growing dendritic actin networks. Such growing networks provide the propulsive force for a 
variety of processes in live cells. A surface growth model is appropriate because polymerization in dendritic actin occurs in a narrow zone next to the load surface, not all over the network. A
distinguishing feature of our model is that it describes the actin network as a growing continuum subject to the balance laws of continuum thermomechanics together with constraints imposed by the dissipation
inequality. The microscopic details of polymerization of individual filaments are distilled into a kinetic law  for the propagation speed of the growing surface. This continuum
model is applied to two different experiments. The first is the set of experiments of Brangbour {\it et al.} \cite{brangbour} in which the density of filaments does not change and the filaments are not cross-linked.
Our model is in remarkable agreement with experiment on the evolution of filament lengths, forces (or stresses), etc., using a maximally dissipative kinetic law. The second is the set of experiments of 
Parekh {\it et al.}\cite{ParekhChaudhuri2005} in which 
the density of filaments changes and the filaments are cross-linked. Once again, a smoothed version of the maximally dissipative kinetic law is able to describe the evolution of variables in this experiment. 
Our kinetic law relates growth rate (in the reference configuration) to an (average) force per filament to facilitate comparison with well-known `force-velocity' relations for single growing filaments. 
We find that the kinetic law that best describes the two experiments is different from the exponential force-velocity relation that is often used in the context of single filaments. This is hardly
surprising since the narrow zone in which polymerization of dendritic actin occurs is a network of interacting filaments which are not parallel and are acted upon by different forces. This kinetic law is one
of many that satisfy the dissipation inequality and there could yet be others that perform better; we use it here because of its simplicity and its ability to capture the convex approach to stall that is
distinct from the well-established exponential force-velocity relation. The continuum framework proposed here may be applicable to growing networks of other biological and non-biological filaments such
as those of collagen, microtubules, carbon-nanotubes, etc.   


\noindent{\bf Acknowledgements} {RA gratefully acknowledges stimulating discussions with Eric Puntel and Giuseppe Tomassetti. This research was partially carried out while PKP was Visiting Professor at 
MIT in Fall 2019.   PKP acknowledges generous support from the Department 
of Mechanical Engineering at MIT and an NIH grant R01 HL 135254.}


\bibliographystyle{abbrv}

\bibliography{BibFile_RA} 


\setcounter{equation}{0}
\renewcommand{\theequation}{A.\arabic{equation}}
\setcounter{figure}{0}
\renewcommand{\thefigure}{A.\arabic{figure}}


\section{Appendix}

\subsection{Dissipation, the driving force and the dissipation inequality.}\label{sec-appendix}
In this section we examine the implications of the dissipation inequality (``second law of thermodynamics'') on the kinetic law for surface growth $V = \mathsf{V}(f)$. Recall that the kinetic laws \eqref{eq-0117-2}$_1$ and \eqref{eq:20191206-3} imply that the growth speed $V$ is $>0$ when the filament force is in the range $0 \leq f <  f_{stall}$.
As we shall see, the dissipation inequality requires $\mathfrak{f}_{driv} V \geq 0$ where $\mathfrak{f}_{driv}(\rho, \sigma)$ is the driving force for growth (which will be identified below). Our task therefore is to examine the implications of the inequality $\mathfrak{f}_{driv} \geq 0$
 on the inequality  $0 \leq f < f_{stall}$, or in terms of stress and filament density, the connection between $\mathfrak{f}_{driv}(\rho, \sigma) \geq 0$ and $0 \leq \sigma < \rho f_{stall}, \, 0 \leq \rho \leq \rho_\infty$.


\subsection{The driving force for growth.}

As shown schematically in Figure \ref{Fig-TestSpecimen.pdf}, the specimen occupies the interval $[y_0(t), y_1(t)]$ in physical space  and its associated length is $\ell = y_1-y_0$. Its left end is attached to an AFM cantilever of stiffness $k_c$. The elastic energy stored in the cantilever (modeled as a Hookean spring) is $\frac 12 k_c (y_0 - Y_0)^2$ where $Y_0$ is the position of the cantilever when it is undeflected. In reference space the specimen occupies the interval $[x_0(t), x_1]$ and its associated length is $\ell_R=x_1 - x_0$.
Let $W$ be the free energy of the specimen per unit reference volume. 
Then, the rate of increase of the energy stored in the specimen and spring is
\begin{equation}
\label{eq-5xx}
  \frac{d}{dt}\big[W A (x_1 - x_0)\big] +   \frac{d}{dt}\Big[ \frac 12 k_c (y_0 - Y_0)^2 \Big]   
  =   \dot W A  \ell_R  - W A \dot x_0 - \sigma A \dot y_0,
\end{equation}
having equated the compressive force  $\sigma A$ in the specimen to the force $-k_c (y_0 - Y_0)$ in the spring. 
The rate at which work is being done on the specimen by the compressive force $\sigma A$  at the 
right-hand end  is 
\begin{equation}
\label{eq:20191010-1}
=-\sigma A \dot y_1.
\end{equation}

Next we model the inflow of chemical energy into the specimen. If $a$ 
denotes the length of a single stress-free monomer (i.e. its length in a 
stress-free reference configuration), the number of monomers in a single  filament is $\ell_R/a$ and the total number of monomers in the specimen is $N \ell_R/a$. Thus the monomer concentration $c$, defined as the number of monomers per unit reference volume, is
$$
 c = \frac{N\ell_R/a}{\ell_RA} = \frac{N}{aA}  = \frac{\rho}{a},
$$
where $\rho = N/A$ is the filament density.  When the body grows,  the left-hand boundary of the specimen moves outwards in reference space at a speed $-\dot x_0$, and so the rate at which monomers are added to a filament at that end is $-\dot x_0/a$. Therefore the associated rate of intake of chemical energy is 
$-\mu \dot x_0/a$ per filament.  In addition, there is an intake of chemical energy due to the formation of new filaments. Since the number of monomers in a filament is $\ell_R/a$ and its chemical energy $\mu \ell_R/a$, the rate of intake of chemical energy due to the formation of new filaments is $(\mu \ell_R/a)\dot N$. 
Thus the total inflow of chemical energy into the specimen per unit time is 
\begin{equation}
 \label{eq-9xx}
-\mu ({\dot x_0}/{a}) N +  (\mu \ell_R/a)\dot N =   
   - \mu c A \dot x_0 + \mu \dot c A \ell_R .
\end{equation}

Therefore the dissipation rate is given by \eqref{eq:20191010-1} plus 
\eqref{eq-9xx} less \eqref{eq-5xx} :
 \begin{equation}
 \label{eq-7xx}
 \begin{array}{lll}
  {\Bbb D} &\displaystyle = -\sigma A \dot y_1 
  + \big[ - \mu c A \dot x_0 + \mu \dot c A \ell_R\big]
 -\big[  \dot W A  \ell_R  - W A \dot x_0 - \sigma A \dot y_0\big] =\\[2ex]
   
  &=\left(-\sigma \lambda  +  \mu c - W \right) A V  
  + \left(- \sigma \dot \lambda +  \mu \, \dot c - \dot W \right) A\ell_R,
\end{array}
\end{equation}
where $\lambda = \ell/\ell_R = (y_1 - y_0)/(x_1 - x_0)$ is the stretch and $V = - \dot x_0$ is the outward propagation speed of the left-hand boundary of the specimen. Suppose that the material is described by the constitutive characterization  $W=W(\lambda, c)$ together with
  \begin{equation}
   \label{eq-11xx}
 - \sigma = \frac{\partial W}{\partial \lambda}, \qquad
\mu = \frac{\partial W}{\partial c},
  \end{equation}
  keeping in mind that $\sigma$ is positive in compression.
 In view of \eqref{eq-11xx},  the dissipation rate \eqref{eq-7xx} reduces to
  \begin{equation}
     \label{eq-12xx}
 {\mathbb D} =  \left(-\sigma \lambda  +  \mu  c- W \right) A V ,
  \end{equation}
 and  we therefore identify the {\it driving force for growth} to be
  \begin{equation}
     \label{eq-13xx}
  \mathfrak{f}_{\rm driv} \coloneqq -\sigma \lambda  +  \mu c - W = -\sigma \lambda  +  \mu \rho/a - W.
  \end{equation}
 Note that $\mu/a$ is the chemical potential per unit reference length.  
By   \eqref{eq-12xx} and   \eqref{eq-13xx}, the dissipation inequality $ {\mathbb D}  \geq 0$ requires
 $$
\mathfrak{f}_{\rm driv} V  \geq 0.
$$
Growth corresponds to $V = - \dot x_0 >0$ which therefore requires $\mathfrak{f}_{\rm driv} \geq 0$.


\subsection{The driving force specialized to the constitutive relation in this paper.}
Next we calculate an explicit expression for the driving force  associated with 
the stress-strain-filament density relation 
\begin{equation}
\label{eq:20200513-1}
 - \sigma = \frac{\partial W}{\partial\lambda} = E\left( 1 - {\lambda}^{-1} \right), \qquad E=E(\rho). 
\end{equation} 
Note from \eqref{eq:20200513-1} and $\lambda >0$ that 
\be
\label{eq:20200514-1}
\sigma > - E.
\ee
Integrating \eqref{eq:20200513-1} with respect to $\lambda$ gives the free energy
\begin{equation}
\label{eq:20200513-2}
 W(\lambda,\rho) = E(\lambda - \log\lambda - 1) + g(\rho), 
\end{equation}
and the corresponding chemical potential is
\begin{equation}
\label{eq:20200513-3}
\mu = \frac{\partial W}{\partial c}  = a\frac{\partial W}{\partial\rho} 
= aE'(\rho)(\lambda - \log\lambda - 1) + ag'(\rho). 
\end{equation}
 The driving force is then given by   \eqref{eq-13xx}, \eqref{eq:20200513-1}, \eqref{eq:20200513-2}  and \eqref{eq:20200513-3} to be
\begin{eqnarray}
 \mathfrak{f}_{\rm driv}  = E(\rho) \log\lambda + \rho E'(\rho)(\lambda - 1 - \log\lambda) + \mathfrak{f}_0(\rho),
\end{eqnarray}
where we have set $\mathfrak{f}_0(\rho) = \rho g'(\rho) - g(\rho)$. This can be written in terms of 
the stress (and filament density) using $\lambda = (1 +\sigma/E)^{-1}$:
\begin{equation}
 \label{eq:inif}
  \mathfrak{f}_{\rm driv} = \mathfrak{f}_{\rm driv}(\rho, \sigma) = -E\log\left(1+\frac{\sigma}{E}\right) + \rho E'
 \left[\frac{1}{1 + {\sigma}/{E}} - 1 + \log\left(1 + \frac{\sigma}{E}
 \right)\right] + \mathfrak{f}_0.   
\end{equation} 
The driving force is a function of both $\sigma$ and $\rho$ since both $E$ and $\mathfrak{f}_0$ depend on $\rho$. All functions of $\rho$ including  $\mathfrak{f}_0(\rho)$ here and $\sigma_{st}(\rho)$ below are defined for $0 \leq \rho \leq \rho_\infty$.

When $\sigma = 0$ (or equivalently $\lambda=1$), the driving force reduces to $ \mathfrak{f}_{\rm driv}  = \mathfrak{f}_0(\rho)$.  This suggests that $\mathfrak{f}_0(\rho)$ is determined entirely
by chemistry and so we refer to it as the {\it chemical driving force}.  
Recall that according to the kinetic relations \eqref{eq-0117-2}$_1$, \eqref{eq:20191111-3} and \eqref{eq:20191206-3}, the growth speed $V$  vanishes when  the filament force $f=f_{stall}$. Since $f=\sigma/\rho$, this means that the growth speed vanishes at the stress $\sigma = f_{stall} \, \rho$.   We now  make the {\it assumption} that the driving force vanishes when the growth-rate vanishes\footnote{but not necessarily the converse. Recall for example from \eqref{eq:20191206-3} that $\mathsf{V}(0) = V_0 >0$.}: 
\be
\label{eq-20200518-1}
 \mathfrak{f}_{\rm driv}(\rho, \sigma_{st}(\rho)) = 0 \qquad {\rm where} \quad \sigma_{st}(\rho) \coloneqq f_{stall} \, \rho \qquad {\rm for} \quad 0 \leq \rho \leq \rho_\infty.
\ee 
Note that $\sigma_{st}(\rho_\infty) = \sigma_{stall}$.  From  \eqref{eq:inif} and \eqref{eq-20200518-1} we now obtain the following expression for the chemical driving force:
\begin{equation}
\label{eq:f0}
 \mathfrak{f}_0(\rho) = E\log\left(1+\frac{\sigma_{st}}{E}\right) - \rho E'
 \left[\frac{1}{1 + \frac{\sigma_{st}}{E}} - 1 + \log(1 + \frac{\sigma_{st}}{E})\right], \qquad   \sigma_{st} = \sigma_{st}(\rho), 
\end{equation} 
where $E=E(\rho)$ as well.
Equation \eqref{eq:f0} can be used to eliminate $\mathfrak{f}_0(\rho)$ in favor of $\sigma_{st}(\rho)$ from  \eqref{eq:inif}  allowing the driving force to be written as
\begin{equation} 
\label{eq:drivf}
 \mathfrak{f}_{\rm driv}(\rho, \sigma) = E\log\left(\frac{1+ \frac{\sigma_{st}}{E}}{1 + \frac{\sigma}{E}} \right)
 + \rho E'\left[\frac{1}{1 + \frac{\sigma}{E}} 
 - \frac{1}{1 + \frac{\sigma_{st}}{E}} - \log \left(\frac{1 + \frac{\sigma_{st}}
 {E}}{1 + \frac{\sigma}{E}}\right)\right]. 
\end{equation}

The driving force  $\mathfrak{f}_{\rm driv}(\rho, \sigma)$ is defined on a subdomain of the $\rho, \sigma$-plane. 
It, and the growth speed $V$, vanish on the bold red line $\sigma = \sigma_{st}(\rho) = f_{stall}\, \rho$  shown in Figure \ref{Fig-rho-sig-plane0}.  The shaded region below that line corresponds to $0 \leq f < f_{stall}$ and therefore to $V >0$ by the kinetic relation. It  remains to examine the consequences of the dissipation inequality $\mathfrak{f}_{\rm driv}(\rho, \sigma) \geq 0$ on this region.


\subsection{The driving force further specialized.}

It order to examine where on the shaded region of Figure \ref{Fig-rho-sig-plane0} one has $\mathfrak{f}_{\rm driv}(\rho, \sigma) \geq 0$ we limit attention to the particular forms of the elastic modulus used in this paper, specifically, $E(\rho) = E_0\rho^n$ where $n=0, 1$ and $2$.

\noindent{\bf Case $E(\rho) = E_0$:}   This case is 
applicable to a specimen involving a few parallel actin filaments and 
the expression  \eqref{eq:drivf} for the driving force specializes  to 
\begin{equation} 
\label{eq:nicex}
 \mathfrak{f}_{\rm driv} = E\log \left(\frac{1 + {\sigma_{st}}/{E}}
                              {1 + {\sigma}/{E}}\right), \qquad \sigma_{st}(\rho) = f_{stall}\rho, \quad E(\rho) = E_0.   
\end{equation}
Keeping \eqref{eq:20200514-1} in mind, we conclude from \eqref{eq:nicex}$_1$ that if $0 < \sigma < \sigma_{st}$ then necessarily $ \mathfrak{f}_{\rm driv}  >0$.  Therefore the dissipation inequality requires $V \geq 0$ on the shaded region of Figure \ref{Fig-rho-sig-plane0} and so all processes obeying the kinetic law $V = \mathsf{V}(f)$ are admissible. In addition, \eqref{eq:nicex}$_1$ shows that $\mathfrak{f}_{\rm driv} $ decreases monotonically with increasing $\sigma$ on this interval, or said differently, the driving force decreases as the stress  becomes progressively more compressive.

\noindent{\bf Case $E(\rho) = \rho E_0$:}   This case pertains to our model of the 
Brangbour {\it et al.} \cite{brangbour} experiments and  
the expression \eqref{eq:drivf}  for the driving force  
reduces to  
\begin{equation}
\label{eq-20200518-5}
 \mathfrak{f}_{\rm driv} = E\left[\frac{1}{1 + \frac{\sigma}{E}} 
 - \frac{1}{1 + \frac{\sigma_{st}}{E}}\right] ,  \qquad \sigma_{st}(\rho) = f_{stall}\, \rho, \quad E(\rho) = E_0\rho. 
\end{equation} 
It follows from \eqref{eq-20200518-5}$_1$ in view of \eqref{eq:20200514-1} that $ \mathfrak{f}_{\rm driv} >0$ when $0 < \sigma <  \sigma_{st}$. Therefore in this case also the dissipation inequality requires $V \geq 0$ on the shaded region of Figure \ref{Fig-rho-sig-plane0} and so all processes satisfying the kinetic law $V = \mathsf{V}(f)$ are admissible on this region.  Moreover, 
$ \mathfrak{f}_{\rm driv}$ decreases monotonically with increasing $\sigma$ (at each fixed $\rho$) as can be seen from \eqref{eq-20200518-5}$_1$, and so the driving force for polymerization decreases as the stress becomes increasingly compressive.

\noindent{\bf Case $E(\rho) = \rho^2 E_0$:}  This case is applicable to the experiments of Parekh {\it et al.} \cite{ParekhChaudhuri2005} and  accounts for filament bending. Recall from \eqref{eq-20200209-2} that in Section \ref{sec-4} we wrote $E = E_\infty \, \rho^2/\rho^2_\infty$, so that in terms of those parameters, $E_0 = E_\infty/\rho^2_\infty$.

\begin{figure}[ht!]
\centering
\includegraphics[scale=0.5]{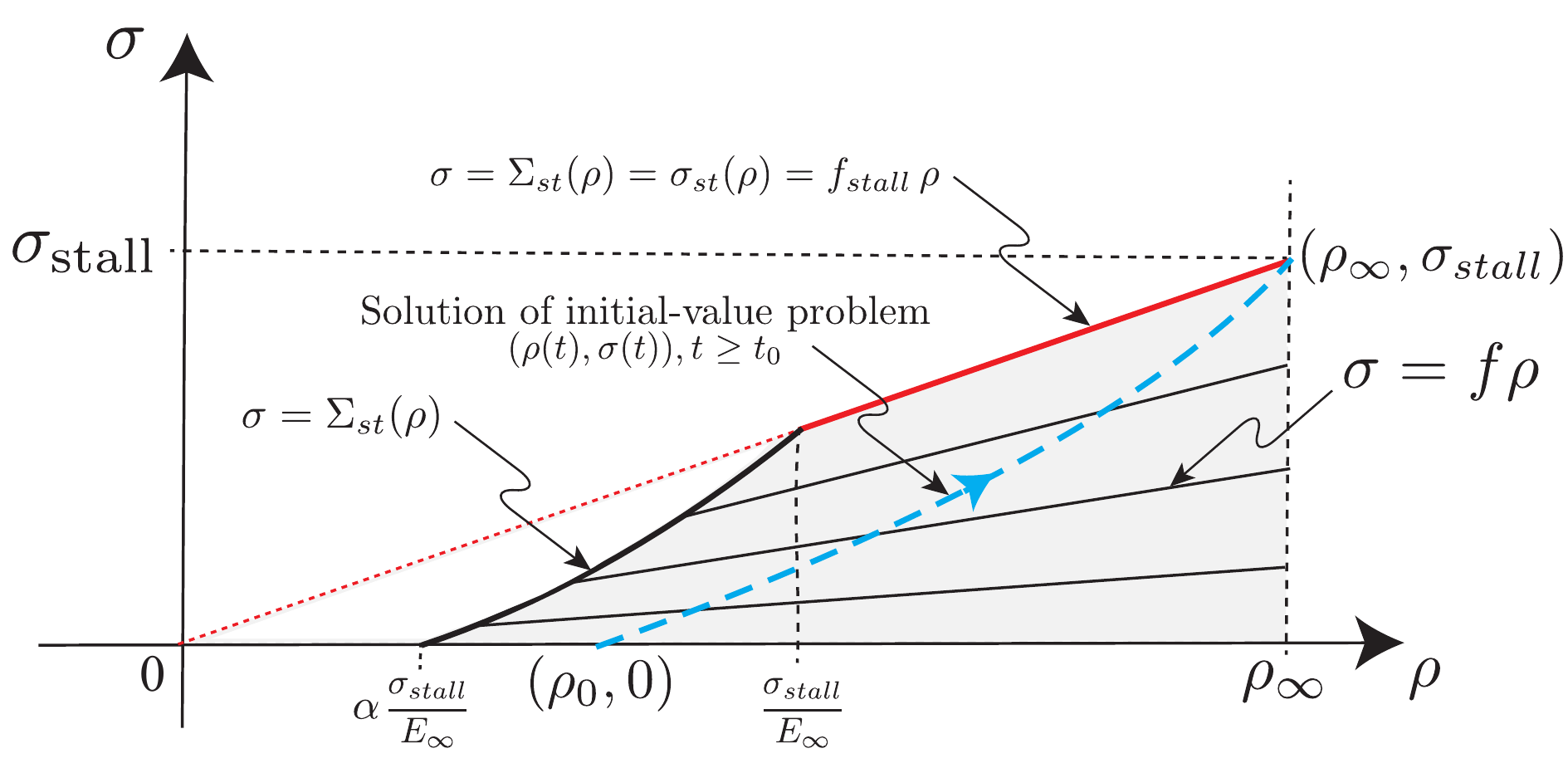}
\caption{Case $E = E_0 \rho^2$: The driving force vanishes on the curve $\sigma = \Sigma_{st}(\rho)$ for $\alpha \sigma_{stall}/E_\infty \leq \rho/\rho_\infty \leq 1$ where $\alpha \approx 0.255$. Note that $\Sigma_{st}(\rho) = \sigma_{st}(\rho)$ for $\sigma_{stall}/E_\infty \leq \rho/\rho_\infty  \leq 1$.  The driving force is positive on the shaded region between the curve $\sigma = \Sigma_{st}(\rho)$ and the $\rho$-axis.  Therefore the dissipation inequality implies that $V \geq 0$ indicating that growth is permitted on this region.  Recall that the kinetic law $V = \mathsf{V}(f)$ also gives $V >0$ here. The trajectory defined by a solution $(\rho(t), \sigma(t)), t \geq t_0,$ of a generic initial-value problem is depicted schematically by the blue dashed curve starting at the initial condition $(\rho_0, 0)$ and terminating at stall corresponding to $(\rho_\infty, \sigma_{stall})$. Only solutions in the shaded region conform to the dissipation inequality.}
\label{Fig-rho-sig-plane2}
\end{figure}

In this case the expression \eqref{eq:drivf} for the driving force specializes to
\begin{equation} 
\label{eq:20200514-9}
 \frac{\mathfrak{f}_{\rm driv}}{E} =   \frac{2}{1 + \frac{\sigma}{E}} 
 - \frac{2}{1 + \frac{\sigma_{st}}{E}} - \log \left(\frac{1 
 + \frac{\sigma_{st}}{E}}{1 + \frac{\sigma}{E}}\right), \qquad \sigma_{st}(\rho) = f_{stall}\rho, \quad E(\rho) = E_\infty\rho^2/\rho_\infty^2.
\end{equation} 
The driving force $\mathfrak{f}_{\rm driv}(\rho, \sigma)$ is defined on the wedge shaped region, $0 \leq \rho \leq \rho_\infty, 0 \leq \sigma \leq \sigma_{st} = f_{stall}\rho$, of the $\rho, \sigma$-plane; see Figure \ref{Fig-rho-sig-plane2}. We want to know where $\mathfrak{f}_{\rm driv}$ is positive on this region.

First consider two limiting cases.  If $\sigma$ is close to $\sigma_{st}$ at fixed $\rho$, one can approximate  \eqref{eq:20200514-9} to read
$$
\frac{\mathfrak{f}_{driv}}{E} = \frac{\left(1 - \frac{\sigma_{stall}}{E_\infty}\, \frac{\rho_\infty}{\rho}\right)}{\left(1 + \frac{\sigma_{st}}{E}\right)^2} \, \frac{\sigma_{st} - \sigma}E \ + \ \mbox{higher order terms},
$$
showing that in this limit the driving force is positive for ${\rho}/{\rho_\infty} > {\sigma_{stall}}/{E_\infty}$ and negative for $ 0 \leq {\rho}/{\rho_\infty} < {\sigma_{stall}}/{E_\infty}$.  Note also that the driving force is proportional to $f_{stall}-f$ in this case. In the other limit where, at fixed $\rho$,  $\sigma$ is small, one gets
$$ 
 \frac{\mathfrak{f}_{\rm driv}}{E} =  2
 - \frac{2}{1 + \frac{\sigma_{st}}{E}} - \log \left({1 
 + \frac{\sigma_{st}}{E}}\right) \ + \ \mbox{higher order terms}.
$$ 
It is not difficult to show that this expression for the driving force  is positive provided $\sigma_{st}/E < 1/\alpha$ or equivalently when $\rho/\rho_\infty > \alpha \, \sigma_{stall}/E_\infty$ where $\alpha \approx 0.255$ is the unique positive root of the equation
$$
2 - \frac{2}{1 + 1/\alpha} - \log \left(1 
 + 1/\alpha\right)  = 0.
$$
Therefore from these two limiting cases we conclude that the driving force is positive on the upper and lower boundaries of the shaded region in Figure \ref{Fig-rho-sig-plane2}.

Returning to the general expression \eqref{eq:20200514-9} for the driving force, keep in mind that we are concerned with the range $0 \leq \sigma/E \leq \sigma_{st}/E $ corresponding to filament force values in the range $0 \leq f \leq f_{stall}$.  In the next paragraph we will show that (at each fixed $\rho$), the equation $\mathfrak{f}_{\rm driv}(\rho, \sigma)=0$ has {\it two} roots $\sigma$, the smaller of which we denote by $\Sigma_{st}$. For $ \sigma_{stall}/E_\infty \leq \rho/\rho_\infty \leq 1$ one has $\Sigma_{st} = \sigma_{st}$, whereas  $\Sigma_{st} < \sigma_{st}$ for $0 \leq \rho/\rho_\infty < \sigma_{stall}/E_\infty$.  Furthermore $\Sigma_{st}>0$ provided $\rho/\rho_\infty > \alpha \, \sigma_{stall}/E_\infty$ where $\alpha \approx  0.255$.  (One cannot write a closed form expression for $\Sigma_{st}$ on this range.)  The curve $\sigma = \Sigma_{st}(\rho)$ for $\alpha \, \sigma_{stall}/E_\infty \leq  \rho/\rho_\infty \leq 1$ is shown in Figure \ref{Fig-rho-sig-plane2}. Finally, we find that 
$\mathfrak{f}_{\rm driv}(\rho, \sigma) > 0$ on the shaded region of the figure between the $\rho$-axis and the curve  $\sigma = \Sigma_{st}(\rho)$.  The figure has been drawn assuming $\sigma_{stall}/E_\infty < 1$. If $\sigma_{stall}/E_\infty > 1$ the curve $\sigma = \Sigma_{st}(\rho)$ lies below the straight line $\sigma = \sigma_{st}(\rho)$ throughout the range of interest.

In order to establish the results described in the preceding paragraph, let $\rho$ be fixed at any value in the interval $(0, \rho_\infty)$, in which event $E(\rho)$ and $\sigma_{st}(\rho)$ are also fixed.
Consider the graph of $\mathfrak{f}_{\rm driv}/E$ versus $\sigma/E$.  First observe that $\mathfrak{f}_{\rm driv}/E \to \infty$ when both $\sigma/E \to -1^+$ and $\sigma/E \to \infty$.  Second, the slope of this curve is negative for $-1 < \sigma/E <1$, positive for  $\sigma/E >1$ and vanishes at $\sigma/E =1$. Third, it intersects the horizontal axis at  $\sigma/E = \sigma_{st}/E$.  It follows that this curve necessarily intersects the horizontal axis at precisely {\it two} points\footnote{except if $\sigma_{st}/E=1$ in which case the two intersection points coalesce.}, corresponding to two values of $\sigma/E \, (>-1)$, one of them less than unity, the other greater.  Let $\sigma/E = \Sigma_{st}/E \leq 1$ correspond to the  left-most intersection point. It then follows that $\mathfrak{f}_{\rm driv} >0$ for $ -1 < \sigma/E < \Sigma_{st}/E$, and  that $\mathfrak{f}_{\rm driv}$ decreases monotonically with increasing $\sigma$ on this interval. It is not difficult to show that $\Sigma_{st} = \sigma_{st}$ if $\rho/\rho_\infty > \sigma_{stall}/E$ and that $\Sigma_{st}/E < 1 < \sigma_{st}/E$ for $0 \leq \rho/\rho_\infty < \sigma_{stall}/E$.

The kinetic law $V = \mathsf{V}(f)$ gives $V > 0$ on the wedge shaped region $0 \leq \rho \leq \rho_\infty, 0 \leq \sigma < \sigma_{st} = f_{stall}\rho$, in Figure  \ref{Fig-rho-sig-plane2}. The dissipation inequality requires $\mathfrak{f}_{driv} V \geq 0$ which reduces to $\mathfrak{f}_{driv} \geq 0$ when $V > 0$. Therefore a solution $\sigma(t), \rho(t), \, t \geq t_0,$ of an initial-value problem involving growth will conform to the dissipation inequality only if the trajectory defined by it lies in the shaded region of  Figure  \ref{Fig-rho-sig-plane2} corresponding to $\mathfrak{f}_{driv} \geq 0$.  We have confirmed this to be true in the case of the particular solutions determined in Section \ref{sec-4} of this paper\footnote{The solutions in Section \ref{sec-3} pertain to the case $E=E_0\rho$, and per the earlier discussion, obey the dissipation inequality.}. The details of this can be found in Section S4 of the  Supplementary Material.

Finally we remark that if the kinetic relation for growth had the form $V = \mathsf{V}(\mathfrak{f}_{driv})$ with $\mathsf{V}(\mathfrak{f}_{driv}) \mathfrak{f}_{driv}\geq 0$ (rather than $V = \mathsf{V}(f)$ with $\mathsf{V}(f)f \geq 0$), all solutions of an initial-value problem would automatically conform with the dissipation inequality.


\clearpage

\renewcommand{\theequation}{\roman{equation}}
\renewcommand{\thesection}{S.\arabic{equation}}
\setcounter{section}{1}

\setcounter{section}{0}
\renewcommand{\thesection}{S.\arabic{equation}}

\setcounter{figure}{0}
\renewcommand{\thefigure}{S.\arabic{figure}}

\begin{center}
{\large \textsc{Online Supplementary Material}}\\
to accompany\\[2ex]

{\bf A Continuum Model for the Growth of Dendritic Actin Networks}\\[2ex]

by\\[2ex]

Rohan Abeyaratne$^{1}$ and Prashant K. Purohit$^{2}$\\[2ex]

$^{1}$Department of Mechanical Engineering, Massachusetts Institute of Technology, Cambridge,  Massachusetts, 02139, USA\\
$^{2}$Department of Mechanical Engineering and Applied Mechanics, University of Pennsylvania,  Philadelphia, Pennsylvania, 19104, USA

\today
\end{center}


\begin{itemize}
\item[\bf S1.]  {\bf Numerical values of the parameters used in modeling the experiments of Brangbour {\it et al.} \cite{xbrangbour}} 
\end{itemize}

From \cite{xbrangbour} we obtain 
\be
V_0 = 0.42\, \si{nm/sec}, \qquad 2R = 1,100\, \si{nm}.
\ee

\begin{itemize}
\item[--] Figure 2 of this paper (Figure 3 of \cite{xbrangbour}):   Brangbour {\it et al.} give $N_{GS} = 4000$ and $c = 0.2 \pm 0.1$. Thus we take $c=0.3$ so that
\be
EA = c \, k_B T \, \frac{N_{GS}}{4R} = 0.3 \times 4.14 \times \frac{4,000}{2200} = 2.258\, \si{pN}.
\ee
They also give the filament length at the end of two stress-free growth periods to be $200 \, \si{nm}$ and $400 \, \si{nm}$. Therefore we take
\be
\ell_R = 200 \, \si{nm} \qquad {\rm and} \qquad  \ell_R = 400 \, \si{nm}.
\ee

\item[--]  Figure 3 of this paper (Figure 1c of \cite{xbrangbour}):  Brangbour {\it et al.} give $N_{GS} = 10,000$. Based on the item above we take $c = 0.2$. Then
\be
EA = c \, k_B T \, \frac{N_{GS}}{4R} = 0.2 \times 4.14 \times \frac{10,000}{2200} = 3.764\, \si{pN}.
\ee

Next we want to calculate the time $t_0$ at which the force was applied.  Let $d(t) = \ell(t) + 2R$; it represents the distance between the centers of two adjacent particles in the model in \cite{xbrangbour}. Then from the formula
\be
\ell(t) = \frac{v_0t}{1+\sigma(t)/E} \qquad \Rightarrow \qquad d(t) =  \frac{v_0t}{1+\sigma(t)/E} + 2R,
\ee
and therefore at the instant $t^+_0$ just after the application of the force, one has
\be
d(t_0^+) =  \frac{v_0t_0}{1+\sigma(t_0^+)/E} + 2R,
\ee
and therefore
\be\label{eq-1}
t_0 = \frac{\big[d(t_0^+) - 2R\big]\big[1 + \sigma(t_0^+)A/EA\big]}{v_0}.
\ee
From Figure 1c of \cite{xbrangbour}, with $EA =  3.764\, \si{pN}, 2R = 1100\, \si{nm}$ and $V_0 = 0.42 \, \si{nm/sec}$, we find

\begin{center}
\begin{tabular}{|c|c|c|}
\hline
$\sigma(t_0^+) A $ & $d(t_0^+)$ & $t_0$\\
 &  & using \eqref{eq-1}\\
\hline
$0.5\, \si{pN}$ & $1460\, \si{nm}$ & $971\, \si{sec}$ \\[2ex]
\hline
$3\, \si{pN}$ & $1320\, \si{nm}$ & $941\, \si{sec}$ \\[2ex]
\hline
$17\, \si{pN}$ & $1180\, \si{nm}$ &$1051\, \si{sec}$ \\[2ex]
\hline
\end{tabular}
\end{center}
Therefore we take $t_0 =1000\, \si{sec}$.

\item[--]  Figure 4 of this paper (Figure 5 of \cite{xbrangbour}): Brangbour {\it et al.} give. $N_{GS} = 4000, \ c=0.13 \pm 0.02$ and $N_{GS} = 10000, \ c=0.18 \pm 0.02$.  So we take $c=0.13$ and $c=0.18$ and then obtain
\be
EA = c \, k_B T \, \frac{N_{GS}}{4R} = 0.13 \times 4.14 \times \frac{4,000}{2200} = 0.979\, \si{pN},
\ee
\be
EA = c \, k_B T \, \frac{N_{GS}}{4R} = 0.18 \times 4.14 \times \frac{10,000}{2200} = 3.387\, \si{pN}.
\ee

\item[--]  Figure 5 of this paper(Figure 2 of \cite{xbrangbour}): Brangbour {\it et al.} give $N_{GS} = 10,000$. Based on the first item above we take $c=0.2$ and then
\be
EA = c \, k_B T \, \frac{N_{GS}}{4R} = 0.2 \times 4.14 \times \frac{10,000}{2200} = 3.764\, \si{pN}
\ee
They also give $t_1 = 650 \, \si{sec}$; $t_2 = 855 \, \si{sec}$; the smaller value of force to be $0.8 \, \si{pN}$; and the larger value of force to be $39 \, \si{pN}$.

\end{itemize}

\vspace{1cm}

\begin{center}
-- CONTINUED --
\end{center}

\clearpage

\begin{itemize}
\item[\bf  S2.] {\bf Numerical values of the parameters used in modeling the experiments of Parekh {\it et al.} \cite{xparekh}}
\end{itemize}

\begin{table}[h]
\scriptsize
\caption{}
\vspace{0.1truein}
\begin{tabular}{|c|c|c|c|l|}
\hline
\multicolumn{5}{|c|}{{\bf Primitive Parameters. From the literature.}} \\[2ex]
\hline
&Quantity &Description & Value & Source\\[2ex]
\hline

1&$k_BT$ &Boltzmann constant times&  $4.14$ \si{pN.nm}& At $\si{300^oK}$\\
&&absolute temperature&&$k_B = 1.381 \times 10^{-23}\, \si{J/K}$\\[2ex]
\hline

2&$E_{\infty}$ & Young's modulus &  $0.7 - 6.7 \, \si{nN/\mu m^2}$& Marcy {\it et al.} \cite{xmarcy}. Average $3.7 \, \si{nN/\mu m^2}$ \\
&&of polymer network&& Parekh {\it et al.} \cite{xparekh} refer to Marcy's data\\
&&&& in their supplement\\[2ex]
\hline

3&$A$ &Specimen cross-sectional area&  $381$ $\si{\mu m^2}$&Parekh {\it et al.} \cite{xparekh} supplementary material\\[2ex]
\hline

4&$k_c$&AFM stiffness & $30\, \si{nN/\mu m}$ &Parekh {\it et al.} \cite{xparekh}   supplement: two cantilevers.\\
&&{\rm  (force/deflection)}& $20\, \si{nN/\mu m}$&$k_c = 0.03 \, \si{nN/nm}$ and $k_c=0.02 \, \si{nN/nm}$\\[2ex]
\hline

5&$\sigma_{stall}A$&Force in specimen at stall&  $294 \,  \si{nN}$&Parekh {\it et al.} 
 \cite{xparekh} Figure 2 \\[2ex]
\hline

6&$\ell_0$ &Length of unstressed specimen& $3000\, \si{nm}$ &Parekh  {\it et al.} 
\cite{xparekh}  Figure 2A. Value of \\
&& at initial instant $t_0$&& $\ell(t)$ at time $t=0$\\[2ex]
\hline 

7&$E_f$&Filament Young's modulus& $2.3\, \si{GPa}$&Howard \cite{xhowardbook}  Table 3.2\\[2ex]
\hline 

8&$A_f$&Filament cross-sectional area& $19\, \si{nm^2} \ [=\pi (2.46)^2]$&Howard \cite{xhowardbook}  Table 7.1\\
&&&&Boal page 24  $\pi (4)^2 = 50\, \si{nm^2}$\\[2ex]
\hline

9&$f_{stall}$&Stall force for one filament& $ 7\, \si{pN}$&Howard \cite{xhowardbook} page 170\\
&&& $8\, \si{pN}$&Ko\v{s}mrlj \cite{xmae545}  \\[2ex]
\hline

10&$a$&Length of a monomer (G-actin)& $2.5\, \si{nm}$&Ko\v{s}mrlj \cite{xmae545}\\
&&&& More or less Howard's $\delta$ \\[2ex]
\hline

\end{tabular}
\label{table-2}
\end{table}


\begin{center}
\begin{table}[h]
\caption{}
\vspace{0.1truein}
\scriptsize
\begin{tabular}{|c|c|c|c|l|}
\hline
\multicolumn{5}{|c|}{{\bf Derived Parameters}}\\[2ex]
\hline
&Quantity &Description & Value & Source\\[2ex]
\hline

a&$\ell_R(0)$ &initial length in ref space & $3000 \, \si{nm}$&$\ell_R(0) =\ell(0)/\mathsf{\Lambda}(0) = \ell(0)$ \\[2ex]
\hline

b&$k$&AFM stiffness &$$ $0.0787\, \si{nN/\mu m^3}$ &$k= k_c/A=30/381$\\
&& {\rm (stress/deflection)}&$$ $0.0525\, \si{nN/\mu m^3}$ &$k= k_c/A=20/381$\\
&&&&\\
\hline

c&$\sigma_{0}$ & $=k\ell_0$ &$$ $ 0.2362\, \si{nN/\mu m^2}$ &$\sigma_{0} = k \ell_0 =0.0787 \times 3$ \\
&&&$$ $ 0.1575\, \si{nN/\mu m^2}$ &$\sigma_{0} = k \ell_0 =0.0525 \times 3$ \\[2ex]

\hline

d&$\sigma_{stall}$&Stress in specimen at stall& $0.77 \, \si{nN/\mu m^2}$&$\sigma_{stall} = 294/A=294/381$\\[2ex]
\hline

&&&&\\
e&$V_0$&$V$ at $f=0$&$32 - 2350 \, \si{nm/min}$&Min value from Brangbour \cite{xbrangbour} \\
&&&&Max value from Marcy {\it et al.} \cite{xmarcy}\\
&&&&See Remark 1\\[2ex]
\hline

f&$\tau_R$ &Time scale for growth &$1.3 - 94\, \si{min}$&$\tau_R=\ell_0/V_0$ with $V_0$ from\\
&&at tips &&row-e\\[2ex]
\hline

g&$\tau_\rho$&Time scale for development &Arbitrary&Unknown\\
&&of new filaments&&\\[2ex]
\hline

\end{tabular}
\label{table-3}
\end{table}
\end{center}

\

\vspace{15cm}
\begin{center}
-- CONTINUED --
\end{center}


\clearpage

\normalsize

\noindent{\bf Remark 1:}  {\bf Estimating $V_0$ in row-e of Table \ref{table-3} from other people's data.}

Brangbour {\it et al.} \cite{xbrangbour},  Figure 5, gives $V = 0.39 \, \si{nm/s}$ at $f = 0.5 \, \si{pN}$. If $V = 
V_0 {\rm exp}(-fa/kT)$, then with $f = 0.5 \, \si{pN}, a = 2.5\, \si{nm}, kT = 4.14 \, \si{pN.nm}$, one gets
$V_0 = 32 \, \si{nm/min}$ (and so $\tau_R =\ell_0/V_0 = 94\, \si{min}$).

Marcy {\it et al.} \cite{xmarcy}, page 5995, right column,  top paragraph gives 
values of $V_0$ (which they call $V_{F = 0}$) in the range $1.75 \pm 0.6 
\, \si{\mu m/min} = 1750 \pm 600 
\, \si{nm/min}$. Therefore $V_0$ can be as large as $2350 
\, \si{nm/min}$. This means $\tau_R = \ell_0/V_0$ can be as small as $1.28\, \si{min}$.

Therefore we have the ranges $V_0 = 32 - 2350 
\, \si{nm/min}$ and $\tau_R = 1.28 - 94\, \si{min}$.

\noindent {\bf Remark 2:}  {\bf Determining which springs were used in the experiments underlying Figure 3 of \cite{xparekh}:} Recall from the supplementary material in Parekh {\it et al.} \cite{xparekh} that they used 2 springs in their experiments with stiffnesses $k_c = 20\, \si{nN/\mu m}$ and $k_c = 30\, \si{nN/\mu m}$.

From the data in Figure 3a of \cite{xparekh} :
$$
\dot\sigma = \frac{145-120}{381\times(100-93.5)} = 0.0100\, \si{nN.min^{-1}.\mu m^{-2}},
$$
$$
k = \frac{\dot\sigma}{\dot\ell} = \frac{0.0100}{129} = 0.0782\, \si{nN/\mu m^{3}}.
$$
By comparing this with the value in the top row of row-b, we infer that they would have used the spring with stiffness $k_c = 30\, \si{nN/\mu m}$ in the experiment related to their Figure 3a.

From the data in Figure 3b of \cite{xparekh}:
$$
\dot\sigma = \frac{83-68}{381\times 6} = 0.00656\, \si{nN.min^{-1}.\mu m^{-2}} ,
$$
$$
k = \frac{0.00656}{129} = 0.0508\, \si{nN/\mu m^{3}}.
$$
By comparing this with the value in the bottom row of row-b, we infer that they would have used the spring with stiffness $k_c = 20\, \si{nN/\mu m}$ in the experiment related to their Figure 3b.

\noindent {\bf Remark 3:}  {\bf Value of $\rho_\infty$.} Though we do not need the value of $\rho_\infty$ since it gets nondimensionalized out, it is still worth calculating it in two different ways, as a consistency check of the model. The first estimate is
$$
\rho_\infty = \frac{\sigma_{stall}}{f_{stall}} = \frac{0.77\, \si{nN/\mu m^2}}{8\, \si{pN}} = 96\, \si{\mu m^{-2}},
$$
and the second follows from $E_\infty = \rho^2_\infty  A^2_f E_f$:
$$
\rho_\infty = \frac{1}{A_f}\sqrt{\frac{E_\infty}{E_f}} =  \frac{1}{19\, \si{nm^2}}\sqrt{\frac{3.7 \si{nN/\mu m^2}}{2.3 \times 10^6\, \si{nN/\mu m^2}}} = 67 \, \si{\mu m^{-2}}.
$$
The upper bound on the number of filaments is
$$
\rho < 1/A_f = 53,000 \, \si{\mu m^{-2}}.
$$

\noindent {\bf Remark 4:}  {\bf Stiffness of the spring used in the experiments underlying Figure 2 of \cite{xparekh}:}
From Figure 2a of \cite{xparekh}
$$
\dot F = \frac{175 -40}{100 - 33} = 2.015 \, \si{nN/min}, \qquad
\dot \ell = \frac{8.5 -4}{100 - 33} = 0.067 \, \si{\mu m/min}, 
$$
and therefore
$$
k_c =\frac{\dot F}{\dot \ell} = \frac{2.015}{0.067} = 30  \, \si{nN/\mu m}.
$$

\vspace{1cm}


\noindent\hrulefill

\noindent{\bf S3.} {\bf Calculations underlying Figures 13 and 14.}

\noindent {\bf S3.1. Figure 13:} Figure 3a of Parekh {\it et al.}  \cite{xparekh} shows that when the specimen grows under spring loading the force increases linearly from the value $120 \, \si{nN}$ to $145 \, \si{nN}$ in $6.5\, \si{min}$. Assuming that the force increased linearly from the start, and extrapolating backwards, we conclude that the force was zero at time\footnote{Presumably prior to that, for $0 < t < t_0$, the actin filaments did not extend all the way from from the AFM cantilever to the other support and so  the specimen was growing freely under stress-free conditions without engaging the AFM spring. } $t_0 = 62.3 \, \si{min}$. Thus we take as initial conditions
$$
\sigma(t_0) = 0, \qquad \rho(t_0) = 0.5 \rho_\infty, \qquad \ell(t_0) = \ell_R(t_0) = 3000\, \si{nm} \qquad \rm{at} \ t_0=  62.3 \, \si{min},
$$
where $0.5 \rho_\infty$ is the arbitrarily chosen initial condition for the filament density and $3000\, \si{nm}$ is the distance between the AFM cantilever and the support when the cantilever is not deflected.
In this experiment the specimen grows under spring loading for $t_0 < t < t_2 = 100\, \si{min}$; at time $t_2$ the stress is suddenly decreased to $0.315\, \si{nN/\mu m^2}$; and thereafter it is held at  $\sigma(t) = 0.315\, \si{nN/\mu m^2}$ for $t > t_2$. The differential equation (4.25) with the preceding initial conditions and $\sigma =k(\ell - \ell_0)$ can now be solved to find $\sigma(t), \dot\ell(t)$ and $\ell(t)$ for $t_0 < t < t_2$. In particular one obtains
$$
\sigma(t_2^-) =0.373\, \si{nN/\mu m^2}, \quad \dot\ell(t_2^-) = 111.6\, \si{nm/min}, \quad \ell(t_2^-) = 7742\, \si{nm}.
$$
Next, the conditions at time $t_2^+$ can be found by first calculating $\ell(t_2^+)$ from (4.27) (keeping in mind that we are given $\sigma(t_2^+)= 0.315\, \si{nN/\mu m^2}$ and we know $\rho(t_2)$ from (4.6)) and $\dot\ell(t_2^+)$ from (4.24):
$$
\sigma(t_2^+) =0.315\, \si{nN/\mu m^2}, \quad \dot\ell(t_2^+) = 212\, \si{nm/min}, \quad \ell(t_2^+) = 8329\, \si{nm}.
$$
 Finally, the differential equation (4.24) is solved for $t > t_2$ using the known information at $t_2^+$ as initial conditions. Figure 13 shows plots of $\sigma(t)$ and $\dot\ell(t)$ versus $t$ resulting from these calculations. This figure is to be compared with Figure 3a of \cite{xparekh}.

\noindent {\bf S3.2: Figure 14.}  As seen in Figure 3b of  \cite{xparekh}, in their second experiment Parekh {\it et al.} kept the stress fixed at the value $\sigma(t) = 0.178\, \si{nN/\mu m^2}$ for $t_0 < t < t_1 = 73 \, \si{min}$; the force clamp was then released at time $t_1$ and the specimen allowed to grow under spring loading conditions for $t_1 < t < t_2 = 79\, \si{min}$; at the instant $t_2$ the value of the stress was  suddenly decreased back to the value $0.178\, \si{nN/\mu m^2}$,  and held there for $t>t_2$.  In order to calculate the response predicted by our model, the first task is to estimate the time $t_0$ at which the stress was initially applied on the specimen (which we assume was done at the instant when the filaments extended all the way from the AFM cantilever to the other support).

In order to determine $t_0$ we proceed as follows: according to Figure 3b of  \cite{xparekh} the stress varies continuously at the instant $t_1$ when the force clamp is released and therefore $\sigma(t_1^+) = 0.178\, \si{nN/\mu m^2}$. Thus from $\sigma(t_1^+) = k (\ell(t_1^+) - \ell_0)$ we find $\ell(t_1^+) = 6390\, \si{nm}$. When $\sigma$ varies continuously it follows from (4.27) that $\ell$ varies continuously (since $\rho$ varies continuously). Thus $\ell(t_1^-) = 6390\, \si{nm}$.  Now focus on the time interval $t_0 < t < t_1$. At the instant $t_0$ we have the initial conditions\footnote{Keep in mind that the length $\ell(t_0^+) \neq 3000\, \si{nm}$ since the specimen length will change suddenly as the stress is applied.}  $\ell_R(t_0^+) = 3000\, \si{nm}$ and  $\rho(t_0) = 0.7 \rho_\infty$ where the initial value of the filament density has been chosen to be consistent with the dissipation inequality; see Section S4.2. Thus integrating $(4.23)_2$ with respect to time from $t_0$ to $t_1$ and using $\ell_R(t_0^+) = 3000\, \si{nm}$ and $\ell_R(t_1^-) = \ell(t_1^-) /\Lambda(\sigma(t_1^-) , \rho(t_1))= 6390/\Lambda(0.178, \rho(t_1))$ with $\rho(t_1)$ given by (4.6) leads to a nonlinear algebraic equation for $t_0$. This yields $t_0 = 48.23\, \si{min}$. Thus we have the initial conditions
$$
\sigma(t_0) = 0.178\, \si{nN/\mu m^2}, \qquad \rho(t_0) = 0.7 \rho_\infty, \qquad \ell_R(t_0) = 3000\, \si{nm} \qquad \rm{at} \ t_0=  48.23 \, \si{min}.
$$

Determining $\sigma(t)$ and $\dot\ell(t)$ is now straightforward.  We first determine $\rho(t)$ for all time from (4.6) with  $\rho(t_0) = 0.7 \rho_\infty$.  We then integrate $(4.23)_2$ with respect to $t$ and use (4.23)$_1$ and the preceding initial conditions to find the stress and elongation-rate for $t_0 < t < t_1$. In particular we obtain 
$$
\sigma(t_1^-) =0.178\, \si{nN/\mu m^2}, \qquad \dot\ell(t_1^-) = 186\, \si{nm/min}, \qquad \ell(t_1^-) = 6390\, \si{nm}.
$$
By the aforementioned continuity of the stress and elongation at time $t_1$ and (4.25) we have
$$
\sigma(t_1^+) =0.178\, \si{nN/\mu m^2}, \qquad \dot\ell(t_1^+) = 124\, \si{nm/min}, \qquad \ell(t_1^+) = 6390\, \si{nm}.
$$
Next we use these as initial conditions to solve (4.25) together with $\sigma = k (\ell - \ell_0)$ to determine  
$\sigma(t)$ and $\dot\ell(t)$  for $t_1 < t < t_2$.  In particular we find
$$
\sigma(t_2^-) =0.216\, \si{nN/\mu m^2}, \qquad \dot\ell(t_2^-) = 119\, \si{nm/min}, \qquad \ell(t_2^-) = 7118\, \si{nm}.
$$
Turning to the instant $t_2^+$, we know the stress $\sigma(t_2^+) = 0.178 \, \si{nN/\mu m^2}$ and so use (4.27) to find $\ell(t_2^+)$. This together with (4.24) gives
$$
\sigma(t_2^+) =0.178\, \si{nN/\mu m^2}, \quad \dot\ell(t_2^+) = 187\, \si{nm/min}, \quad \ell(t_2^+) = 7508\, \si{nm}.
$$
Finally we solve (4.24) to find the stress and elongation-rate for $t >t_2$ using the above information at $t_2^+$ as initial conditions.
Figure 14 shows plots of $\sigma(t)$ and $\dot\ell(t)$ versus $t$ resulting from these calculations. This figure is to be compared with Figure 3b of \cite{xparekh}.

\vspace{1cm}

\noindent\hrulefill

\noindent{\bf S4.} {\bf The solutions in Sections 4.2 and 4.3 obey the dissipation inequality.}

\noindent{\bf S4.1.} {\bf Spring-loaded solution.}

\begin{figure}[h]
\centerline{\includegraphics[scale=0.3]{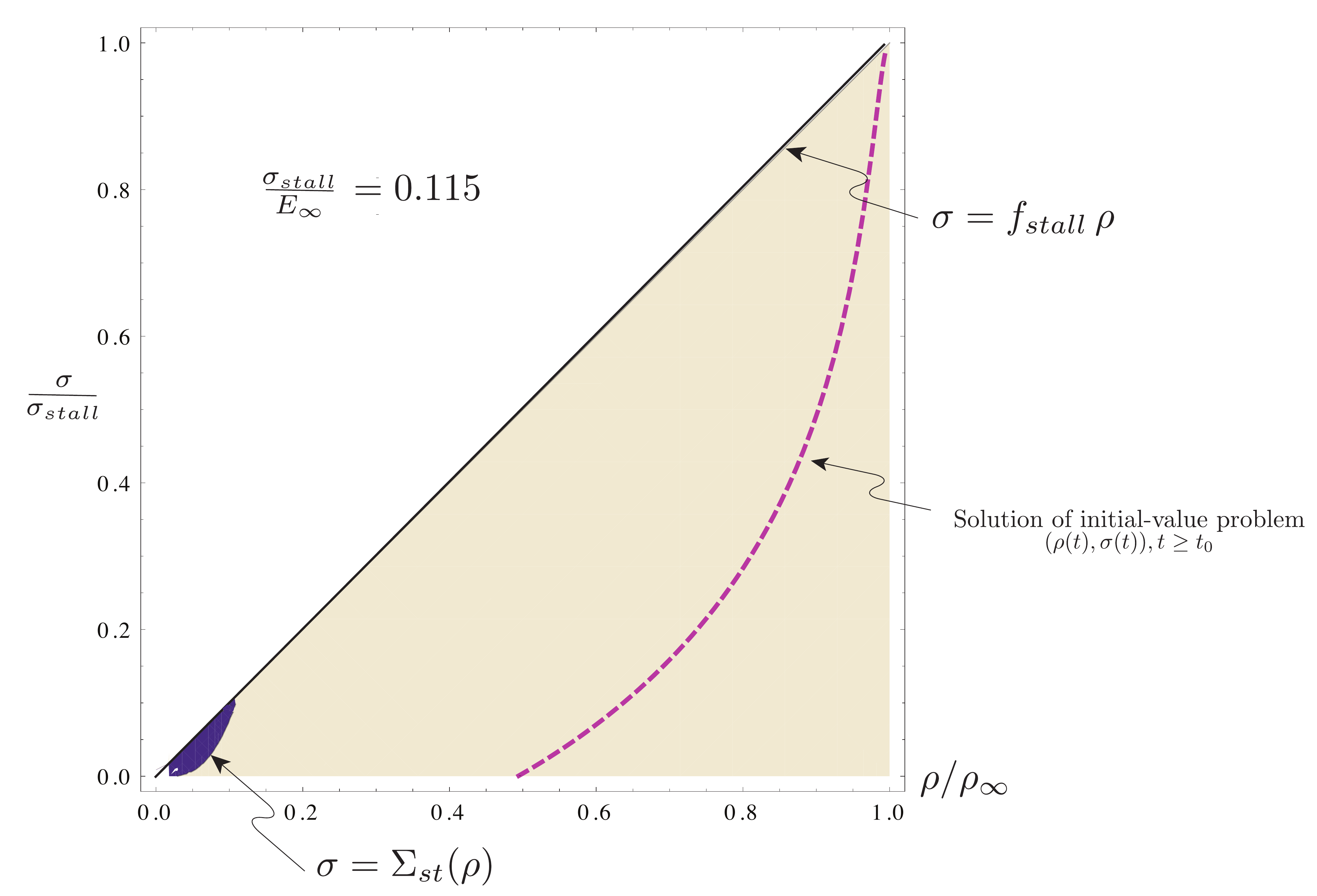}}
\caption{{\footnotesize{The specimen is spring-loaded per Section 4.2. The dissipation inequality and kinetic relation both give $V >0$ in the lightly shaded region. The reader is referred to the
appendix for a definition of the stress $\Sigma_{st}(\rho)$.}}}
\label{Fig-ZeroDF-Spring}
\end{figure}

The solution $\rho(t), \sigma(t)$, $t \geq t_0$, found in Section $4.2.2$ for the spring loaded specimen is shown  on the  $\rho, \sigma$-plane in  Figure \ref{Fig-ZeroDF-Spring}. The trajectory starts from $(\rho(t_0), \sigma(t_0)) = (0.5\rho_\infty, 0)$ and terminates at stall corresponding to $(\rho_\infty, \sigma_{stall})$. Since $\sigma_{stall} = 0.77\, \si{nN/\mu m^2}$ and $E_\infty = 3.7 \, \si{nN/\mu m^2}$, the figure has been drawn with  $\sigma_{stall} /E_\infty = 0.115$.  We are only concerned with the region on and below the straight line $\sigma = f_{stall} \rho$ since the filament force then lies in the range $0 \leq f \leq f_{stall}$. The driving force is positive in the lightly shaded region below this line, and therefore on this region both the dissipation inequality $\mathfrak{f}_{driv}\, V \geq 0$ and the kinetic relation $V = \mathsf{V}(f)$ give $V >0$. The solution associated with the trajectory shown therefore satisfies the dissipation inequality.
%


\begin{itemize}
\item[\bf S4.2.] {\bf Loading programs involving discontinuous stress.}
\end{itemize}

Next consider the loading programs studied in Section $4.3$ involving a discontinuous change in stress at a certain instant $t_2$. Since $\sigma_{stall} = 0.77 \, \si{nN/\mu m^2}$ and $E_\infty = 0.7 \, \si{nN/\mu m^2}$, in this case we have  $\sigma_{stall} /E_\infty = 1.1$.

\begin{figure}[h]
\centerline{\includegraphics[scale=0.3]{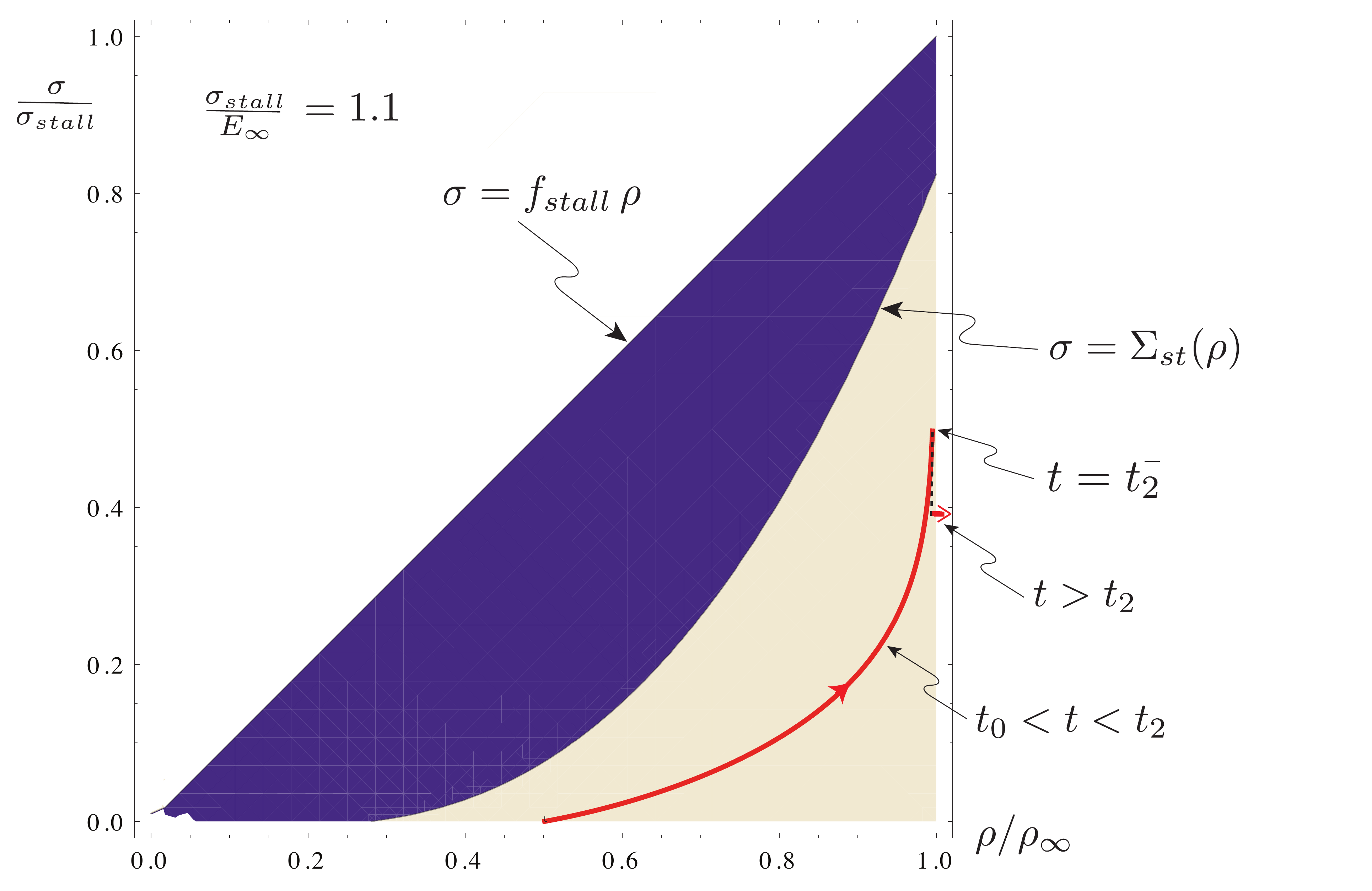}}

\caption{{\footnotesize The specimen is spring-loaded for $t_0 < t < t_2$. The stress is decreased suddenly at time $t_2$ and then held constant for $t > t_2$. Here $t_0 = 0$ and $t_2 =100\, \si{min}$. The
reader is referred to the appendix for the definition of the stress $\Sigma_{st}(\rho)$. }}

\label{Fig-ZeroDF-Jump1}
\end{figure}

In the first calculation, the specimen initially grows under spring loading conditions and follows the trajectory shown in  Figure \ref{Fig-ZeroDF-Jump1} for $t_0 < t <t_2$. It starts from $(\rho(t_0), \sigma(t_0)) = (0.5\rho_\infty, 0)$ and evolves towards stall.  However, when the stress reaches the value $0.49\sigma_{stall} $ at time $t_2$, it is suddenly decreased to the value $0.41\sigma_{stall} $ and held constant at that value for $t > t_2$. The trajectory therefore drops vertically at this instant and follows a rightward pointing horizontal line thereafter; this part of the trajectory falls within a very narrow sliver and is barely visible in Figure \ref{Fig-ZeroDF-Jump1}.  Again,  the trajectory lies in the lightly shaded part of the figure where $\mathfrak{f}_{driv}>0$ and $ V >0$ and therefore the solution satisfies the dissipation inequality.

\begin{figure}[h]
\centerline{\includegraphics[scale=0.3]{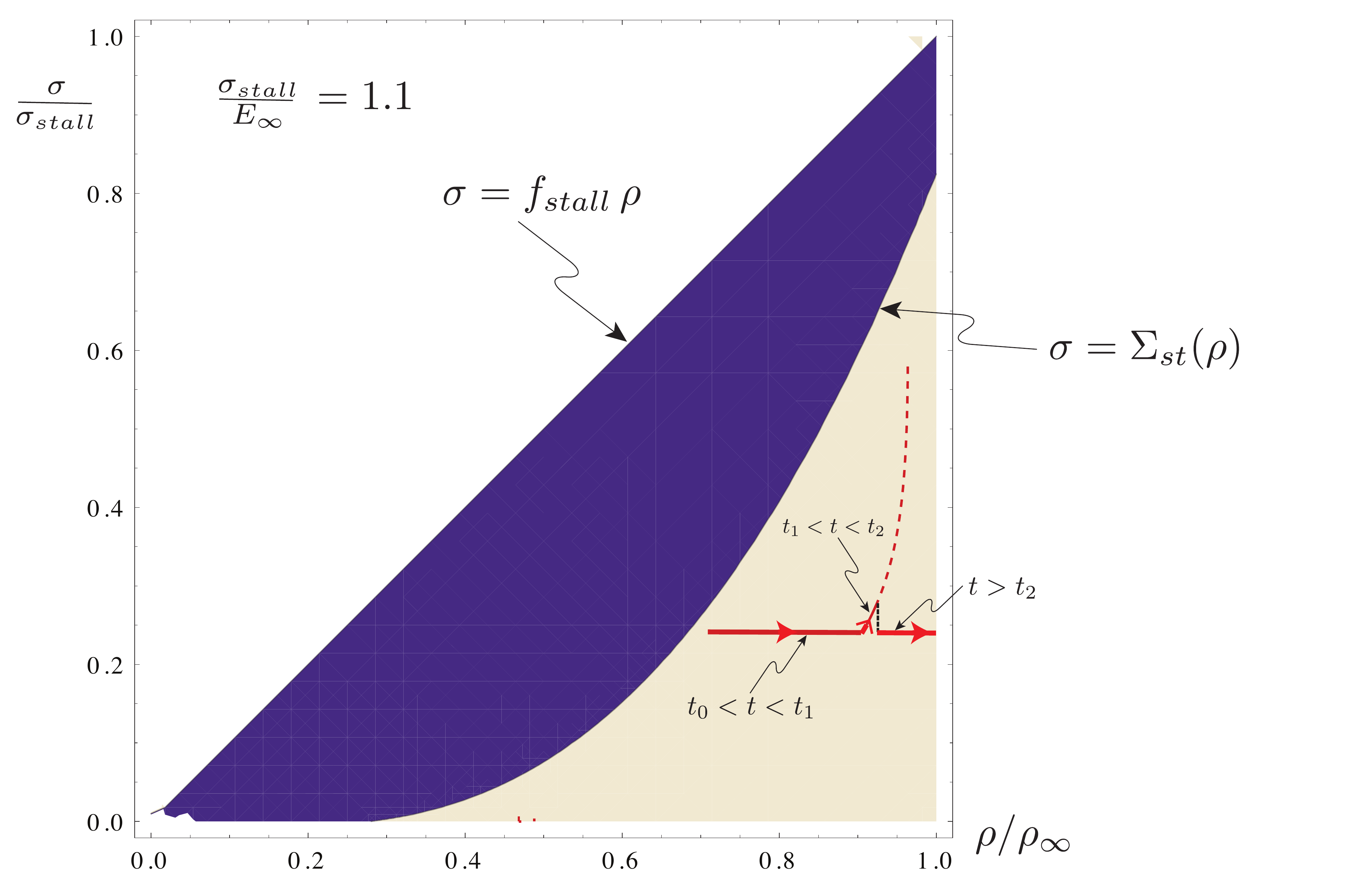}}

\caption{{\footnotesize{The stress is kept constant at the value $0.23 \sigma_{stall} $ for time $t_0 < t < t_1$. The force clamp is released at time $t_1$ and the specimen is spring-loaded for $t_1 < t < t_2$. At time $t_2$ the stress is decreased suddenly back to its original value, and kept constant for $t > t_2$.  Here $t_0 = 0, t_1 = 73\, \si{min} $ and $t_2 = 79\, \si{min}$. The reader is referred to the appendix for the definition of the stress $\Sigma_{st}(\rho)$.}}}

\label{Fig-ZeroDF-Jump2}
\end{figure}

In the second calculation, initially, for a period $t_0 \leq t \leq t_1$, the stress  is held constant  at the value  $0.23 \sigma_{stall} $ and so the trajectory of $(\rho(t), \sigma(t))$ follows a rightward pointing horizontal line on the $\rho, \sigma$-plane.  In order for this trajectory to be admissible by the dissipation inequality it must lie in the lightly shaded region in Figure \ref{Fig-ZeroDF-Jump2} and therefore $\rho(t_0)$ must exceed the value $\approx 0.6765$.  We take $\rho(t_0) =0.7 \rho_\infty$. At the instant $t_1$ the force clamp is released and the specimen now grows under spring loading conditions. The trajectory for $t_1 < t < t_2$ is determined as in Section $4.2$ with the initial condition $\sigma(t_1)/\sigma_{stall} = 0.23$, $\rho(t_1)/\rho_\infty = 0.919$ where the value of $\rho(t_1)$ was determined using equation (24). The solution  now follows the curved trajectory shown in Figure \ref{Fig-ZeroDF-Jump2}.  At time $t_2$, when the stress has reached the value $0.28 \sigma_{stall} $, it is suddenly decreased back to $0.23 \sigma_{stall}$ and held constant at that value from then on.   The trajectory lies in the lightly shaded region of the figure and therefore satisfies the dissipation inequality.

%


\end{document}